\documentclass[final,5p,times,twocolumn]{elsarticle}
\usepackage{graphicx}
\usepackage{mathptmx}      
\usepackage{latexsym}
\usepackage[T1]{fontenc}       
\usepackage{graphicx}
\usepackage{bm}
\usepackage{amsmath,amsfonts,amssymb}
\usepackage{amssymb}
\usepackage{natbib}

\begin{document}

\begin{frontmatter}

\title{Deuteration of ammonia with D atoms on oxidized partly ASW covered graphite surface}.




\author{Henda Chaabouni\corref{cor1}\fnref{label1}} \author{Marco Minissale\corref{cor2}\fnref{label2}} \author{Saoud Baouche\fnref{label1}} \author{Francois Dulieu\fnref{label1}}

\cortext[cor1]{Henda Chaabouni; henda.chaabouni@u-cergy.fr}

\address{$^1$ Universit\'{e} de Cergy Pontoise, 5 mail Gay Lussac, 95031 Cergy Pontoise Cedex, France.\\
              LERMA, UMR 8112 CNRS, Observatoire de Paris, Sorbonne Universit\'{e}, UPMC Univ. Paris 6, PSL Research University.\fnref{label1}}
\fntext[label1]{Universit\'{e} de Cergy Pontoise, 5 mail Gay Lussac, 95031 Cergy Pontoise.}

\cortext[cor2]{Marco Minissale; marco.minissale@univ-amu.fr}

\address{$^2$ Aix Marseille Universit\'e, CNRS, PIIM UMR 7345, 13397 Marseille, France.\\
                            Aix-Marseille Universit\'e, CNRS, Centrale Marseille, Institut Fresnel UMR 7249, 13013 Marseille, France.\fnref{label2}}
\fntext[label2]{Aix Marseille Universit\'e, CNRS, PIIM UMR 7345, 13397 Marseille.}


\begin{abstract}

The deuteration of ammonia by D atoms has been investigated experimentally in the sub-monolayer regime on realistic analogues of interstellar dust grain surfaces. About 0.8~monolayer of solid ${\rm NH_3}$ was deposited on top of an oxidized graphite surface held at 10~K, partly covered with ASW ice. Ammonia ice is subsequently exposed to D atoms for different exposure times using a differentially pumped beam-line. The deuteration experiments of ammonia were carried out by mass spectroscopy and temperature programmed desorption (TPD) technique.
The experimental results showed the formation of three  isotopologue ammonia species by direct exothermic H-D substitution
surface reactions: ${\rm NH_3+D}$~$\rightarrow$~${\rm NH_2D+H}$,  ${\rm NH_2D +D}$~$\rightarrow$~${\rm NHD_2+H}$, and
${\rm NHD_2+D}$~$\rightarrow$~${\rm ND_3+H}$.
The formation of the deuterated isotopologues ${\rm NH_2D}$, ${\rm NHD_2}$, and ${\rm ND_3}$ at low surface temperature (10~K) is likely to occur through quantum tunneling process on the oxidized graphite surface.
A kinetic model taking into account the diffusion of D atoms on the surface is developed to estimate the width and the hight of the activation energy barriers for the successive deuteration reactions of ammonia species by D atoms. Identical control experiments were performed using CH$_3$OH and D~atoms. The deuteration process of solid methanol is ruled by H~abstraction and D addition mechanism, and is almost five orders of magnitude faster than ammonia deuteration process.

\end{abstract}

\end{frontmatter}



\section{Introduction}

Ammonia molecules are detected in the gas phase of molecular clouds: Taurus Molecular Cloud-1 (TMC1-N)
\citep{1968Cheung}, with relatively high abundances 10$^{-7}$-10$^{-8}$ respective to H$_2$ molecules \citep{Hama2013}. Solid NH$_3$ has been detected through infrared absorption in different astrophysical environments: high-mass
protostars \citep{2004Gibb}, low-mass protostars \citep{2010Bottinelli}, comets \citep{2004BockeleeMorvan}, and in dense molecular
clouds \citep{2001Dartois}. The interstellar grain mantles in dense molecular clouds are predominantly composed of ${\rm H_2O}$ ice, combined with
other molecules such as (CO, ${\rm CO_2}$, ${\rm NH_3}$, ${\rm H_2CO}$, and ${\rm CH_3OH}$) \cite{1982Tielens,2010Bottinelli}.
The abundance of ammonia (NH$_3$) in the icy mantles is 1 to 15~\% with respect to water (H$_2$O) ice
\citep{2000Gibb,2004Gibb}, while in the cold dust envelopes of young stellar objects, the ammonia ice fraction is 5~\% or less
\citep{2001Dartois}. In comets, ammonia is present at the 1~\% level
relative to water ice \citep{2004BockeleeMorvan}.
Deuterated ammonia NH$_2$D was first detected by Rodriguez Kuiper et al~\citep{1978Rodriguez} in high temperature molecular clouds such as
Orion-KL Nebula region (T=50-150~K).
NH$_2$D molecules have been also observed in many sources towards dark molecular clouds \citep{2000Saito},
galactic protostellar cores~\citep{2001Shah}, and interstellar dense cores (L134N) \citep{2000Tine}. The [NH$_2$D]/[NH$_3$]
ratio in gas phase varies from 0.02 to 0.1. These abundance ratios are larger than the cosmic abundance of elemental deuterium relative to hydrogen (D/H),
which is expected from the Big-Bang nucleosynthesis to be 1.5$\times$10$^{-5}$~\citep{2003Linsky}.
Observations in low mass protostellar cores showed the highest [NH$_2$D]/[NH$_3$] ratios (0.3), indicating that deuterium fractionation of ammonia increases towards protostellar regions \citep{2003Hatchell}. Chemical models explained this high fractionation ratio by the gas-phase
ion-molecule chemistry with depletion of C, O and CO from the gas phase \citep{2000Tine,2003Hatchell}.
Doubly deuterated ND$_2$H ammonia has been also detected for the first time in cold, 10~K, dense cores L134N by~Roueff et al.~\citep{2000Roueff}.
The expected fractionation ratio [ND$_2$H]/[NH$_3$] from models is 0.03 \citep{2003Hatchell}.
The Caltech Submillimeter Observatory (CSO) has detected the triply deuterated ammonia, ND$_3$, through its J$_K$ emission
transition near 310~GHz \citep{2002Van} in cold clouds (10~K). The observed [ND$_3$]/[NH$_3$] ratio in very cold clouds of gas and dust is found to be closer to 0.001.
Such a high isotopic ratio between ND$_3$ and NH$_3$ suggested that the deuteration of NH$_3$ is likely to occur by ion-molecule reactions in the gas phase, in which deuteron transfer reactions are much faster than proton transfer \citep{2002Van,2002Lis}.
Theoretical models of pure gas-phase chemistry \citep{2000Robert,2000Saito,2005Roueff} explained relatively well the abundances of simply
and multiply deuterated ammonia molecules in dense cores. According to Tielens et al.~\citep{1983Tielens}, grain surface chemistry would also build deuterated
molecules by deuteration process on grain mantles with D atoms. The trapped deuterated species on grains are eventually released into the gas phase due to the heating of a close star in the formation stage. Recent chemical models of cold dark clouds \cite{2004Roberts} have shown that desorption of species into the gas phase via thermal evaporation is negligible for dark clouds with temperatures of 10~K.

The deuteration experiments of solid ammonia by D atoms has been already performed by two astrophysical groups; the Watanabe group (Nagaoka et al. \cite{2005Nagaoka}), and the Leiden group (Fedoseev et al. \cite{2015Fedoseev} using mainly infrared spectroscopy. The experimental studies of Nagaoka et al.~\cite{2005Nagaoka} have shown an efficient deuteration of CH$_3$OH ice by D atoms addition at low surface temperature. The deuterated methanol species are formed via H-abstraction and D-addition mechanism, and through tunneling quantum reactions. These authors have reported that no deuterated species of ammonia are observed in the exposure of pre-deposited NH$_3$ ice to D atoms at temperatures below 15~K. Even the experimental results of Fedoseev et al. \cite{2015Fedoseev} have also shown that the deuteration of solid NH$_3$ by D atoms did not take place at temperature lower than 15~K, by depositing D atoms on ammonia ice, or by performing co-deposition experiments of NH$_3$ molecules with D atoms on gold cold surface.

Based on the previous experimental results of Fedoseev et al. \citep{2015Fedoseev} and Nagaoka et al. \citep{2005Nagaoka}, someone wonders about the dramatic difference observed in the deuteration of ammonia and methanol by D atoms in the solid phase. If these authors \citep{2015Fedoseev,2005Nagaoka} did not observe the deuteration of the NH$_3$ by D atoms in their experiments, this is probably because of the very high activation energy barrier of the reaction NH$_3$+D in the solid phase in comparison to that of ${\rm CH_3OH}$+D. The value of the activation energy barrier of the reaction (NH$_3$~+~D~$\rightarrow$~NH$_2$D~+~H) has been estimated from earlier experimental \citep{1969Kurylo} and theoretical \citep{2005Moyano} works in the gas phase to be ${\rm 11~kcal\cdot mol^{-1}}$ or ${\rm 46~kJ\cdot mol^{-1}}$. While the activation energy barrier of the abstraction reaction (${\rm CH_3OH~+~D}$~${\longrightarrow}$~${\rm CH_2OH +HD}$), has been reported from gas phase estimations to be lower than that of ammonia (${\rm 27~kJ\cdot mol^{-1}}$) \cite{Hama2013}. But up to now, there is no laboratory studies providing activation energy barriers for the reaction ${\rm CH_3OH+D}$ and that of ${\rm NH_3+D}$ in the solid phase.

It is obvious that laboratory experiments are important for understanding the deuteration reactions occurring on the cold grain surfaces between condensed molecules and the impinging D atoms. However, some factors related to the gas flux of the deuterium atoms, the thickness of the ices on the grains, and the fluences of the atomic species on the surface, may affect the progress and the evolution of these reactions.
In the case of the the previous works of Nagaoka et al. \citep{2005Nagaoka}, and Fedoseev et al. \citep{2015Fedoseev}, the authors have performed experiments in the multi-layer regime by covering the aluminium surface with 10~ML of solid NH$_3$ \citep{2005Nagaoka}, and the the gold surface by 50~ML of ammonia ice \citep{2015Fedoseev}, and irradiating the corresponding ices with D-flux of 1-4$\times$10$^{13}$ ${\rm atoms\cdot cm^{-2}\cdot s^{-1}}$, and 3.7$\times$10$^{14}$~${\rm atoms\cdot cm^{-2}\cdot s^{-1}}$, respectively (see Table~\ref{tablea}).
First of all, the use of a high flux of D atoms in their experiments favors the recombination reactions D+D on the surface or in the bulk of the ices, and reduces therefore the reaction efficiency of D atoms with the adsorbed ${\rm CH_3OH}$ and ${\rm NH_3}$ species. However, in the experiments of Nagaoka et al. \citep{2005Nagaoka}, the deuteration reaction (${\rm CH_3OH+D}$) seems to be not affected by the high flux of D atoms. This is probably because the activation energy barrier of H-D exchange reaction between D and ${\rm CH_3OH}$ is lower that that between D and ${\rm NH_3}$.
On the other hand, as reported by Fedoseev et al. \citep{2015Fedoseev}, the use of a thick layer of ammonia ice favors the formation of hydrogen bonds (N-H), that can strength the interaction ${\rm NH_3}$-${\rm NH_3}$ molecules, and prevent the D-H exchange between D atoms and adsorbed ${\rm NH_3}$ molecules.

In this work, we performed deuteration experiments of solid ${\rm NH_3}$ by D atoms in the sub-monolayer regime, and with low D-flux, on an oxidized highly oriented pyrolytic graphite (HOPG) surface, partly covered with an amorphous solid water (ASW) ice held at 10~K of less than 0.5~ML of thickness, simulating water ice contaminations.
We deposited only a fraction of one monolayer of solid ammonia (0.8~ML) on the substrate to study the effect of the grain surface on the efficiency of the deuteration reaction between D atoms and the adsorbed ${\rm NH_3}$ molecules. In this work, we considered that the physisorption of species on the oxidized HOPG surface dominates the chemisorption process.
We also used low D-flux in comparison to the previous works \citep{2015Fedoseev,2005Nagaoka} to reduce the recombination efficiency of D atoms on the surface, and increase the probability of the H-D substitution reaction. As shown in the Table~\ref{tablea}, at our experimental conditions, even by reducing the D atoms Fluence by factors 100 and 10, with respect to those of Nagaoka et al. \cite{2005Nagaoka} and Fedoseev et al. \cite{2015Fedoseev}, respectively, the total amount of D atoms (53.5~ML) sent on the surface seems to be sufficient for the D-fractionation of solid ammonia, and the formation of the deuterated species ${\rm NH_2D}$, ${\rm NHD_2}$, and ${\rm ND_3}$.

For comparison, similar D atoms addition experiments have been performed with ${\rm CH_3OH}$ molecules to corroborate the findings of Nagaoka et al.~\citep{2007Nagaoka}, and validate the deuteration method governed by the abstraction-addition mechanism. D atom addition and H atom abstraction may not be the only mechanism to deuterate molecules on ices. Direct H-D substitution reactions could also proceed at low temperatures to fractionate the astrophysical molecules.

\begin{table*}
\centering \caption{Comparison between the experimental conditions and the results of the system (${\rm NH_3+D}$) for different works and references. \label{tablea}}
\begin{tabular}{ccccccc}
\hline\hline
 Article                                   &  D Fluence   &  D Thickness     & D-Flux       & ${\rm NH_3}$  Thickness        & results  \\
             &              &          &   \\
\hline
                                            & ${\rm atoms\cdot cm^{-2}}$ & ML & ${\rm atoms\cdot cm^{-2}\cdot s{-1}}$  &  ML                  &  \\
\hline
Nagaoka et al. \cite{2005Nagaoka} &  $\leq$10$^{18}$  & $\leq$1000    &10$^{14}$                      & 10      & no deuteration \\
(2015)                                                    &              &                           &   \\
Fedoseev et al. \cite{2015Fedoseev}  &  8 $\times$ 8$^{16}$-3$\times$ 10$^{17}$ & $\leq$100   &1-4$\times$ 10$^{13}$     & 50                  &  no deuteration \\
(2015)                                                    &              &                           &   \\
This work  & $\leq$5.35$\times$ 10$^{16}$  & $\leq$53.5       &3.7 $\times$ 10$^{12}$     & 0.8                  &  deuteration \\
\hline\hline
\end{tabular}
\end{table*}

The paper is organized as follows: in section~\ref{exp}, we describe the experimental setup and explain the procedures of the deuteration experiments; section~\ref{results} presents the experimental results for ${\rm NH_3~+~D}$ and ${\rm CH_3OH~+~D}$ reactions, and in the section~\ref{model and discussion}, we propose a kinetic model to estimate the activation energy barriers of the successive H-D substitution reactions. We make some concluding remarks in the final section.

\section{Experimental}\label{exp}

The experiments were performed with the FORMOLISM (FORmation of MOLecules in the InterStellar Medium)
apparatus.
The experimental setup is briefly described here and more details are given in a previous work \citep{2007Amiaud}.
The apparatus is composed of an ultra-high vacuum (UHV) stainless steel chamber with a base pressure of about $10^{-11}$ mbar.
The sample holder is located in the center of the main chamber. It is thermally connected to a cold finger of a closed-cycle Helium cryostat.
The temperature of the sample is measured in the range of 6~K-350~K.  The sample holder is made of a 1~cm diameter copper block which is covered with a highly
oriented pyrolytic graphite (HOPG, ZYA-grade) substrate. The HOPG is a model of an ordered carbonaceous material mimicking  interstellar dust grains analogues in astrophysical environments. It is characterized by an arrangement
of carbon atoms in a hexagonal lattice. The HOPG grade (10~mm diameter $\times$ 2~mm thickness) was firstly dried in an oven at about 100~$^{\circ}$C during two hours, and then cleaved several times using "Scotch tape" method at room temperature to yield several large terraces (micron scale) that contain limited defects and step edges. The HOPG was cleaved in air immediately prior to being inserted into the vacuum chamber. It was mounted directly onto the copper finger by means of a glue (ARMECO Product INC CERAMA BOND 571-P).
In chamber, the HOPG sample was annealed to 300~K under UHV to remove any contaminants. In this work, we used an oxidized HOPG sample, which has been preliminary exposed to oxygen atomic beam under UHV for several exposure doses, and then warmed-up from 10~K to 300~K to desorb oxygen and other species from the substrate, mainly water molecules. The oxidation phase was achieved after the saturation bonds of the surface, the defects and the step edges of the sample. This behavior was deduced when there is no modification in the Thermally Programmed Desorption profiles of the adsorbates. Prior oxidation of the HOPG is expected to give stable surface, where the structure cannot be modified by other adsorbates.

FORMOLISM is equipped with a quadrupole mass spectrometer (QMS) which allows routinely the simultaneous detection
of several species in the gas phase by their masses. The QMS can be placed either in front of the surface for the detection of species desorbed into the gas phase during the warming-up of the sample, or in front of the beam-line for the characterization and the calibration of the NH$_3$, CH$_3$OH, and D atoms beams. The experimental setup is also equipped with a Fourier transform infrared spectrometer (FTIR) for the in-situ solid phase measurements by reflection absorption infrared spectroscopy (RAIRS) in the spectral range 4000-700 (${\rm cm^{-1}}$ \cite{2012Chaabouni}.

The D atomic jet is prepared in a triply differentially pumped beam-line aimed at the sample holder. Its is composed of three vacuum chambers connected togethers by tight diaphragms of 3~mm diameters. The beam-line is equipped with a quartz tube with inner diameter of 4~mm, which is surrounded by a microwave source cavity for the dissociation of D$_2$ molecules. When the microwave source (Sairem) is turned on, the cavity is cooled down with a pressurized air jet, and D atoms are produced from the D$_2$ molecular plasma. The D$_2$ plasma is generated by a microwave power supply coupled into a Surfatron cavity operating at 2.45~GHz and producing up to 300~W. The warm D atoms undergo several collisions with the inner walls of the tube, and finally thermalize at the room temperature of about 350~K before they reach the surface . However, the charged particles composed of exited atoms, ions and electrons, produced in the plasma quickly recombine within the tight quartz tube \cite{2010Ioppolo, 2011Theule}. Because of the high micro-wave frequency, the hot energetic particles cannot leave the discharge pipe, as reported in some astrophysical laboratory works \cite{2010Ioppolo}.
The deuterium beam dissociation rate, measured with the quadrupole mass spectrometer from the ${\rm D_2}$ signals (m/z=4) during the discharge (ON) and the discharge (OFF) of the microwave source is calculated from the following relationship ${\rm \tau=\frac{D_2(OFF)-D_2(ON)}{D_2(OFF)}}$. In this work, the dissociation rate ${\rm \tau}$ of ${\rm D_2}$ beam reaches a high value of 85~\% with an effective microwave power of 50~W.

The flux of the dissociated D atoms coming from the gas phase and hitting the surface is
${\rm \Phi_{D, ON}}$ = ${\rm (3.7 \pm 0.5)}$ ${\rm \times10^{12}}$ ${\rm atom\cdot cm^{-2}\cdot s^{-1}}$. It is defined as ${\rm \Phi_{D, ON}}$ = ${\rm 2\tau~\Phi_{D_2,Off}}$,
where ${\rm \Phi_{D_2,Off}}$ = ${\rm (2.2 \pm 0.4)}$ ${\rm \times10^{12}}$ ${\rm molecule\cdot cm^{-2}\cdot s^{-1}}$ is the
flux of D$_2$ beam before running the microwave discharge.
The flux of D$_2$ molecules coming from the beam-line is determined by the so-called the King and Wells method \citep{1972King}, which is generally used to evaluate the sticking coefficient of particles incident on a cold surface. This method consists to measure with the mass spectrometry the indirect D$_2$ signal in the vacuum chamber in the real time during the exposure of D$_2$ on the oxidized HOPG surface.
${\rm \Phi_{D_2}}$ is calculated from the ratio between the exposure dose of D$_2$ molecules that saturate the graphitic surface, expressed in (${\rm molecule\cdot cm^{-2}}$) by the corresponding exposure time of D$_2$ on the surface, expressed in (second). According to the estimation made by Amiaud et al. \cite{2007Amiaud}, a compact ice layer begins to saturate after an exposure to 0.45~ML of D$_2$ (i.e. ${\rm 0.45\times10^{15}}$ ${\rm molecule\cdot cm^{-2}}$). More detail description and errors about the estimation of D-flux are given in the reference \cite{2007Amiaud}.

In this study, we have deposited all species (NH$_3$, CH$_3$OH, H$_2$O molecules, and D atoms) by using only one beam-line, oriented at 45$^{\circ}$ relatively to the surface of the sample. That guarantees a quasi-perfect match between the effective areas on which particles are deposited.

\begin{figure}
\centering
\includegraphics[width=8cm]{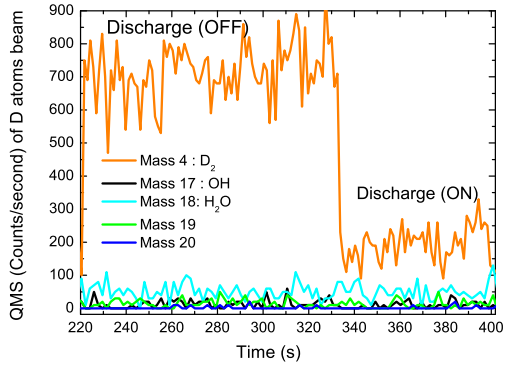}
\caption{The QMS signals (in counts/seconds) as a function of the time (s) of m/z=2 (D), m/z=4 (D$_2$), m/z=17 (NH$_3$), m/z=18 (NH$_2$D, ND$_2$H), m/z=19 (ND$_2$H), and m/z=20 (ND$_3$), given by the QMS, placed in front of the the D beam after the deposition of the NH$_3$ molecules on the oxidized HOPG surface using the same beamline.}
\label{Fig8}
\end{figure}

In our NH$_3$+D experiments, the beam-line is pumped off to evacuate the residual gas of ammonia species, after the deposition phase of ${\rm NH_3}$ ice on the cold surface. Then the D atoms are generated in the same beam-line by the microwave dissociation of D$_2$ molecules. We have checked with the QMS, placed in front of the beam-line, that no deuterated species (${\rm NH_{2}D}$, ${\rm ND_2H}$, ${\rm HDO}$, ${\rm ND_3}$, ${\rm D_{2}O}$) and radicals (ND$_2$, OD, OH) contaminants are coming from the D beam. Figure~\ref{Fig8} shows the signal of (m/z=4) before the dissociation of deuterium molecules (discharge OFF) and during the dissociation phase (discharge ON). We notice that there is no increase in the signal of mass 18 during the discharge ON, which may correspond to ${\rm NH_{2}D}$ and ${\rm ND_2}$ species formed from NH$_3$ and D atoms within  the beam-line. The small signal of  mass 18 is the background signal of
${\rm H_2O}$ molecules contaminants in the main chamber.  Moreover, the absence of the signals (m/z=19) and (m/z=20) excludes any possible formation of ${\rm NHD_{2}}$ and ${\rm ND_3}$ species in the D beam.

Ammonia and methanol ice films, with a thickness of 0.8~ML
were grown on the oxidized HOPG surface held at 10~K by beam-line vapor deposition
of NH$_3$ molecules (from Eurisotop bottle with 99.9~\% purity) and CH$_3$OH molecules (from liquid methanol with 99.5~\% purity).
The monolayer surface coverage corresponds to the number density of molecules that populate ${\rm 10^{15}}$ sites on the surface. It is defined as ${\rm 1~ML}$ = ${\rm 10^{15}}$ ${\rm molecules\cdot cm^{-2}}$.
In this work, the fluxes of ammonia and methanol species that hit the surface, are defined as the amounts of these species that saturate the surface per unit time ${\rm (\Phi}$ ${\rm =\frac{exposure~dose~for~1~ML}{exposure~time~for~1~ML}}$). The values of the fluxes are found to be ${\rm \Phi_{NH_3}}$ ${\rm =2.1\times10^{12}}$  ${\rm molecules\cdot cm^{-2}\cdot s^{-1}}$
and ${\rm \Phi_{CH_{3}OH}}$  ${\rm = 1.7\times10^{12}}$ ${\rm molecules\cdot cm^{-2}\cdot s^{-1}}$.

Because water is always present as contaminant in the ultrahigh vacuum chamber, and can be condensed on the cold surface at 10~K, we have performed experiments to study the effect of the water ice on the deuteration of solid ammonia. In order to simulate the small amount of the water ice that can be condensed on the surface during the exposure phase of the reactants at 10~K, we have deposited a very thin film of porous amorphous solid water (ASW) ice with $\sim$0.5~ML of thickness on the oxidized graphite surface at 10~K, by H$_2$O vapor deposition during 5 minutes, using the same beam-line as that for ammonia and D atoms. The water vapor was obtained from deionized water which had been purified by several pumping cycles under cryogenic vacuum.

We have estimated the thickness in ML of the amorphous water ice film grown on the surface at 10~K by beam-line vapor (${\rm H_2O}$) deposition using the reflection absorption infrared spectroscopy (RAIRS). We have deposited water ice on the surface at 10~K, for different exposure times, then we have recorded the RAIR spectra, and measured the integrated areas $\int \nu d\nu$ in (${\rm cm^{-1}}$) below the IR absorption bands of water ice at about 3200~${\rm cm^{-1}}$.  Using the formulae  ${\rm S=\frac{Ln10 \int \nu d\nu} {N}}$ \cite{2007Bisschop}, where N is the column density of water ice in (molecule $\cdot$ cm$^{-2}$), and S is the strength band of ${\rm H_2O}$ at 3200~${\rm cm^{-1}}$ in (cm~$\cdot$ molecule$^{-1}$), we have estimated the exposure time required to form 1~ML surface coverage of ASW ice at 10~K. The first monolayer of amorphous water ice covering the surface at 10~K is reached after about 10~minutes of water deposition time with a flux of ${\rm 10^{12} molecules\cdot cm^{-2}\cdot s^{-1}}$. The absorbance value of the corresponding ${\rm H_2O}$ infrared band at 3200 ${\rm cm^{-1}}$ is very low (AB=0.0005). Because of the low surface coverage of solid ${\rm NH_3}$ (0.8 ML) on oxidized graphite, and the low absorbance (AB=0.0002) of the infrared band of ammonia at 1106 ${\rm cm^{-2}}$ for the vibrational mode ${\rm \nu_2}$, the deuteration experiments of ammonia by D atoms on the oxidized graphite surface has been analyzed in this work only by TPD-QMS spectroscopic method.

For the deuteration experiments of solid NH$_3$ (or CH$_3$OH), we firstly deposited 0.8~ML of NH$_3$ (or CH$_3$OH) ices
on the HOPG surface held at 10~K, and then we exposed the films of ammonia (or methanol) to D atomic beam at the same surface temperature.
After the exposure phases, we used TPD technique by warming-up
the sample from 10~K to 210~K with a linear heating rate of ${\rm 0.17~K\cdot s^{-1}}$, until the
sublimation of the ices from the surface. The species desorbed into the gas phase are then detected and identified through mass spectrometry.

\section{Results}\label{results}

\subsection{Co-deposition of H$_{2}$O-NH$_{3}$}\label{H2O-NH3}

\begin{figure}
\centering
\includegraphics[width=8cm]{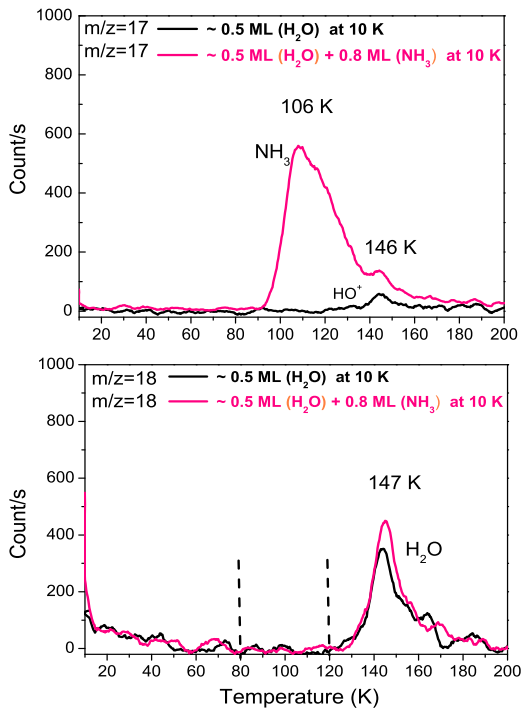}
\caption{TPD signals of mass m/z=17 (Top panel) and mass m/z=18 (bottom panel) between 10~K and 200~K. Black curve: ($\sim$~0.5)~ML of pure porous amorphous water (${\rm H_2O}$) ice deposited on the oxidized graphite surface at 10~K. Magenta curve: ${\rm ~0.8~ML}$ of NH$_3$ ice deposited on top of ($\sim$~0.5)~ML of ${\rm H_2O}$ ice pre-deposited on the oxidized graphite surface at 10~K.} \label{Fig1}
\end{figure}

In order to study the effect of the water on the adsorption-desorption of ammonia molecules, two kinds of experiments have been performed on the graphite surface held at 10~K using ${\rm H_2O}$ and NH$_3$ molecules.
In the first experiment, we grow (${\rm \sim 0.5~ML}$) of amorphous solid water (ASW) ice on the oxidized graphite surface by exposing the sample held at 10~K to ${\rm H_2O}$ water beam during 5 minutes. The TPD curve of the water ice desorbed from the surface during the heating phase is displayed in the Figure~\ref{Fig1}, bottom panel for mass (m/z= 18) between 10~K and 200~K. The maximum of this desorption peak is centered at about 147~K.
In the second experiment, we deposited 0.8~ML of solid NH$_3$ on top of 0.5~ML surface coverage of water (${\rm H_2O}$) ice grown on the oxidized graphite surface at 10~K.
Figure~\ref{Fig1}, top panel shows the signals of m/z=17 after the warming-up of pure ${\rm~H_2O}$ and (${\rm~H_2O-NH_3}$) ices from 10~K to 200~K. The small TPD peaks (in black and magenta lines) at 147~K correspond to the cracking pattern ${\rm OH^+}$ (m/z=17) of ionized ${\rm H_2O^+}$ molecules desorbing from the oxidized graphite surface. While the TPD peak (in magenta line) (m/z=17) at 106~K corresponds to the desorption of pure ionized NH$_3$ molecules from the surface.

Figure~\ref{Fig1}, bottom panel compares the TPD signals of m/z=18 before and after the addition of solid ammonia on top of ${\rm H_2O}$ ice. For the two ${\rm~H_2O}$ and (${\rm~H_2O-NH_3}$) ices, the Figure~\ref{Fig1}, bottom panel shows  only one desorption peak at around 147~K, which is slightly more intense for ${\rm~H_2O-NH_3}$ ice (in magenta line) than pure ${\rm H_2O}$ ice (in black line). The small increase in the area below the desorption curve of m/z=18 at 147~K for the  ${\rm~H_2O-NH_3}$ ice, may result either from the instability of the ${\rm H_2O}$ flux, or from the desorption of the ${\rm NH_4^{+}}$ (m/z=18) compound formed by the protonation reaction between NH$_3$ and  ${\rm H_2O}$ molecules, as has been observed by Souda \cite{2016Souda} in its recent work for the interaction of  NH$_3$ with porous  ASW ice.

\subsection{Exposure of NH$_{3}$ and D atoms on graphite surface}\label{NH3+D}

\begin{figure*}
\centering
\includegraphics[width=18cm]{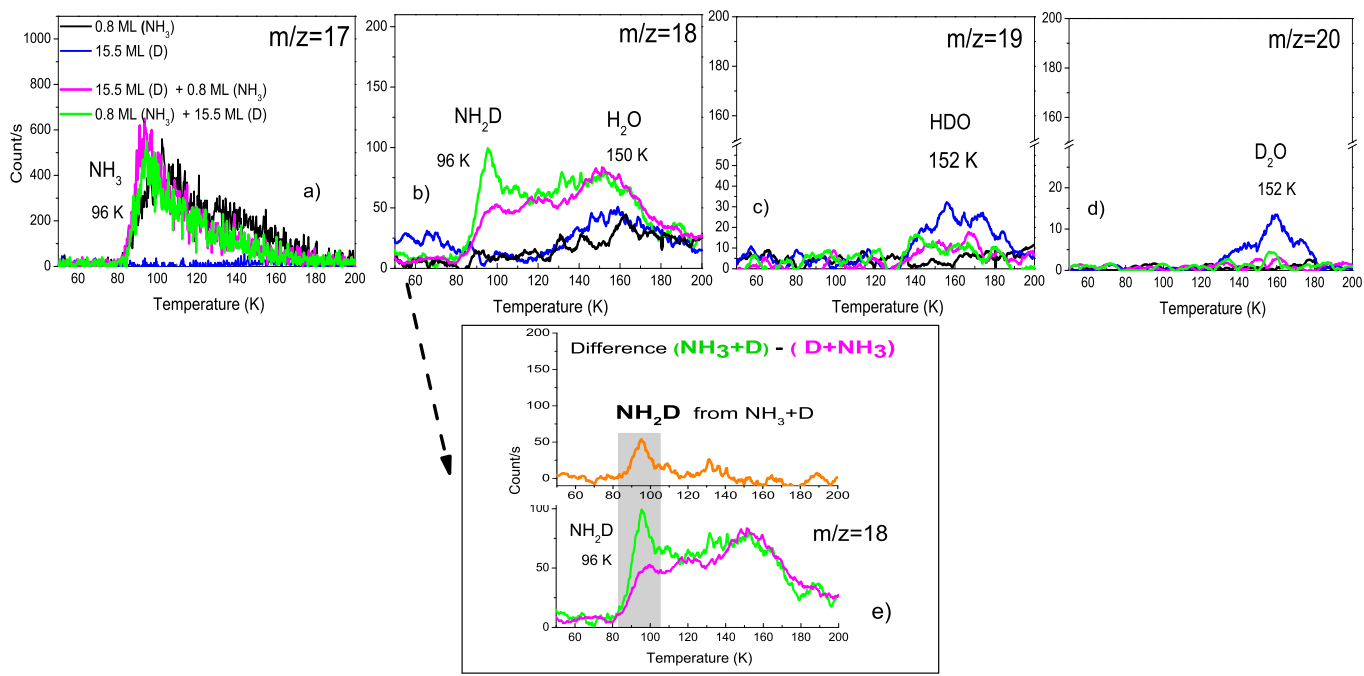}
\caption{TPD signals between 50~K and 200~K of masses: a) m/z=17, b) m/z=18, c) m/z=19, and d) m/z=20. Black curve: deposition of 0.8~ML of NH$_3$ ice, blue curve: deposition of 15.5~ML of D atoms, magenta curve: deposition of the film 1 (15.5~ML~D~atoms~+~0.8~ML~NH$_3$), green curve: deposition of the film 2 (0.8~ML~NH$_3$~+~15.5~ML D atoms). The deposition of the species is performed on an oxidized HOPG surface held at 10~K. The Figure e) gives the difference between the green curve (Film 2) and the magenta curve (Film 1) for mass 18. It illustrates the desorption peak of NH$_2$D (m/z=18) at around 96~K, really formed by the surface reaction NH$_3$~+~D.} \label{Fig2}
\end{figure*}

In the first experiment, we prepared a film~1 (${\rm 15.5}$~ML~(D)~+~${\rm 0.8~ML}$~(NH$_3$)), by exposing firstly the oxidized HOPG surface at 10~K, to D beam for ${\rm 15.5 ML}$ surface coverage, and then to 0.8~ML of NH$_3$ ice at the same surface temperature 10~K. In the second experiment, the film~2 (${\rm 0.8~ML}$ (NH$_3$)~+~${\rm 15.5~ML}$ (D)), is prepared by exposing ${\rm 15.5~ML}$ of D atoms on top of 0.8~ML of NH$_3$ ice pre-deposited on the oxidized HOPG surface at 10~K.
Two TPD control experiments were also performed in addition to the previous ones, by depositing separately 0.8~ML of solid NH$_3$, and ${\rm 15.5~ML}$ of D atoms on the oxidized HOPG surface. The TPD curves of all the experiments are displayed in the Figure~\ref{Fig3} for the masses 17, 18, 19 and 20.
The TPD curves of the two films of NH$_3$ (0.8~ML) and that of D atoms (${\rm 15.5~ML}$) deposited separately on the surface, showed small peaks at around 152~K, for masses 18, 19 and 20 (see panels (b), (c) and (d) of Figure~\ref{Fig2}). These peaks correspond to the desorption of water contaminations, such as H$_2$O (m/z=18), HDO (m/z=19), and D$_2$O (m/z=20). These water impurities came either from the beam-line during the exposure phases of NH$_3$ and D atoms, or from the ultra-high vacuum chamber.
In the case of the film~1, where D atoms are deposited before NH$_3$ on the surface, the TPD curve of mass 18 in the  Figure~\ref{Fig2}b shows a small desorption peak at around 96~K overlapping a large peak at 130~K-150~K. While in the panels (c) and (d) of the Figure~\ref{Fig2}, we observe only the desorption peaks at 152~K for the masses 19 and 20, respectively. The desorption peaks at the higher temperatures 150~K and 152~K for the masses 18, 19 and 20 are likely to be originated from water impurities H$_2$O, HDO and D$_2$O, respectively.
Based on the computational results of Burke et al. \cite{1983Burke}, the sticking coefficients of the impinging D atoms coming from the gas phase at room temperature onto graphite and ASW ice (held at 10~K) is 90~\% and  60~\%, respectively. However, the experimental studies of Matar et al. \citep{2010Matar} for the sticking coefficient of D atoms on the non-porous ASW ice is estimated to be 30~\%. Since our substrate is constituted of an oxidized HOPG, partly coved with ASW ice contaminants (H$_2$O, HDO and D$_2$O), most of the D atoms exposed on the substrate for the (film~1), will stick both on the oxidized graphite and on the water surface adsorption sites. These atoms promptly form D$_2$ molecules by D~+~D surface recombination, either by Langmuir-Hinshelwood mechanism based on the diffusion of two adsorbed D atoms on the surface, or via Eley-Rideal abstraction reactions between adsorbed D atoms and incoming D atoms from the gas phase \citep{2002Zecho,2005Cazaux}. Moreover, the experimental and the theoretical studies of Horneker et al. \cite{2006Horneker} have revealed a possible route for D$_2$ formation on the HOPG surface through D adsorbate clusters. The D$_2$ molecules formed on the graphitic surface cannot react with the NH$_3$ molecules adsorbed on the surface, and cannot be therefore involved in the formation of the new isotopic species of ammonia.
We suggested that water contaminants present on the surface, such as HDO, D$_2$O, may react with the deposited NH$_3$ molecules and form NH$_2$D species (m/z=18) through the following exothermic reactions (\ref{Eq1a}) and  (\ref{Eq1b}), provided by Nist web-book~\citep{Nist}.

\begin{equation}\label{Eq1a}
{\rm NH_3+HDO~\rightarrow~NH_2D~+~H_2O},
\end{equation}
\begin{equation}\label{Eq1b}
{\rm NH_3+D_2O~\rightarrow~NH_2D~+~HDO},
\end{equation}

The presence of a very small desorption peak (m/z=20) for the deuterated water at 150~K following exposure to ammonia (magenta traces in Figure~\ref{Fig2}d) may support our suggestion. The NH$_2$D molecules that can be formed by isotopic exchange reaction between ${\rm NH_3}$ molecules and HDO and D$_2$O species on the oxidized graphite surface may desorb between 50~K and 120~K.
Moreover, the exposure of D atoms on the oxidized graphite surface may create new functional groups or intermediates, such as (-OD). These reactive intermediate species may interact with ${\rm NH_3}$ and form NH$_2$D following the exothermic reaction (\ref{Eq1c}), provided by Nist web-book~\citep{Nist}.

\begin{equation}\label{Eq1c}
{\rm NH_3~+~OD~\rightarrow~NH_2D~+~OH},
\end{equation}\label{Eq1c}

All these suggestions for the formation of NH$_2$D by heavy water contaminants or by -OD intermediates, could explain the observed small desorption peak (in magenta), for mass 18, in the Figure~\ref{Fig2}b, at around 96~K (film~1), where NH$_3$ molecules are deposited on top of D atoms on the oxidized graphite surface.
In the case of the film~2 (${\rm 0.8~ML~NH_3}$+${\rm 15.5 ML}$~of~D), where D atoms are deposited on top of the solid NH$_3$ film, the desorption peak (in green) at 96~K for mass 18 becomes larger than previously (see Figure~\ref{Fig2}b). The increase in the TPD area of the peak at 96~K for mass 18 is likely due from the reaction of D atoms with NH$_3$ molecules on the surface. The TPD peak at 96~K of the NH$_2$D molecules produced only from the NH$_3$~+~D reaction on the oxidized HOPG substrate with water deposits, is shown in the Figure~\ref{Fig2}e. The desorption curve (in orange) is the difference (film~2-film~1), between the TPD signal of NH$_2$D (m/z=18), expected to be formed on the surface in the film 2, by the TPD signal of NH$_2$D (m/z=18), produced by the reaction of NH$_3$ with -OD, HDO and/or D$_2$O contaminants on the surface in the film~1.

\subsection{Kinetics of NH$_3$+D reaction}\label{Kinetic NH3+D}

\begin{figure*}
\centering
\includegraphics[width=18cm]{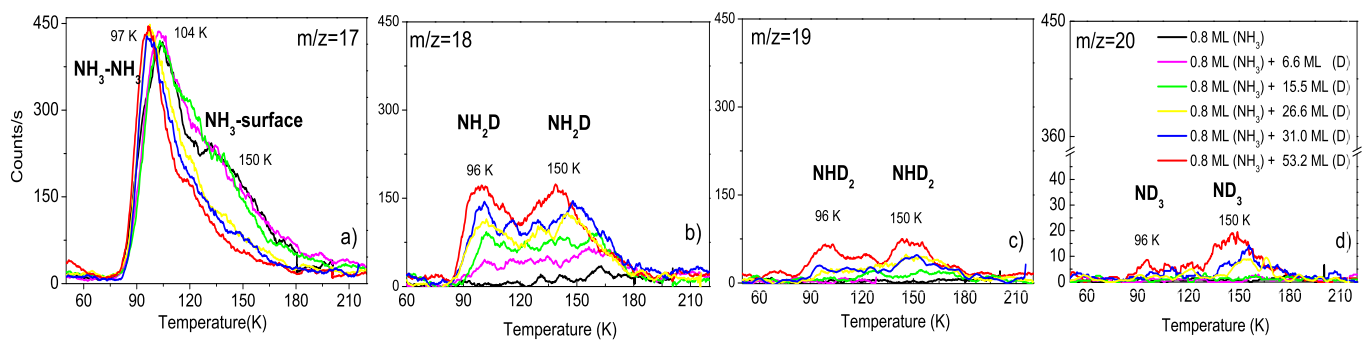}
\caption{TPD desorption curves of ammonia species between 60~K and 220~K as a function of D-atoms exposure doses (0 ML, ${\rm 6.6~ML}$, ${\rm 15.5~ML}$, ${\rm 26.6~ML}$ ${\rm 31.0~ML}$ and ${\rm 53.2~ML}$) on 0.8~ML of solid NH$_3$ ice grown on the oxidized HOPG surface held at 10~K with amorphous water ice contaminants: a) m/z=17 (${\rm NH_3}$), b) m/z=18 (${\rm NH_2D}$), c) m/z=19 (${\rm NHD_2}$), and d) m/z=20 (${\rm ND_3}$).}
\label{Fig3}
\end{figure*}

The kinetic reaction ${\rm NH_3+D}$ has been studied by exposing 0.8~ML of solid NH$_3$ to different doses of D atoms
(0 ML, ${\rm 1.0~ML}$, ${\rm 6.6~ML}$, ${\rm 15.5~ML}$, ${\rm 31.0~ML}$ and ${\rm 53.2~ML}$).
TPD curves of species for masses 17, 18, 19 and 20 are shown in Figure~\ref{Fig3} between 50~K and 210~K, for each film of ${\rm NH_3}$ and D atoms. As shown in the Figure~\ref{Fig3}a, the maximum of the TPD peak of NH$_3$ shifts slightly toward the higher temperatures with the D-exposure time from 96~K to 104~K, and in parallel we observe the disappearance of a second desorption peak, as a shoulder at about 150~K. These desorption temperatures differences can be explained in terms of reaction sites and/or surface contamination, such as water molecules. In these experiments the amount of water ices contaminants on the surface  is negligible.

In addition, panels (b), (c) and (d) of Figure~\ref{Fig3} show the growth of three double desorption peaks at 96~K and 150~K, for masses 18, 19 and 20, respectively.
The desorption peak at about 96~K in Figure~\ref{Fig3}b is likely to be attributed to NH$_2$D (m/z=18) species, mainly produced from the reaction between ${\rm NH_3}$ and D atoms. Similarly, the desorption peaks observed at 96~K in Figure~\ref{Fig3}c, and d are attributed to the doubly deuterated species ${\rm NHD_2}$ (m/z=19), and triply deuterated ammonia ${\rm ND_3}$ (m/z=20), formed mainly by the reaction  NH$_2$D+D, and  NHD$_2$+D, respectively. We neglected the contribution of NH$_2$D, ${\rm NHD_2}$ ${\rm ND_3}$ formed from the contaminants on the surface, mainly water ices in these experiments.
Furthermore, the cracking pattern of the ionized  ammonia molecules ${\rm ND_3^{+}}$  (m/z=20), ${\rm NHD_2^{+}}$ (m/z=19),  ${\rm NH_2D^{+}}$ (m/z=18), and  ${\rm NH_3^{+}}$  (m/z=17) by electron impact, in the ion source of the QMS, are  ${\rm ND_2^{+}}$, ${\rm NHD^{+}}$, ${\rm NH_2^{+}}$, ${\rm ND^{+}}$, ${\rm NH^{+}}$, ${\rm D_2^{+}}$, ${\rm H_2^{+}}$, ${\rm D^{+}}$, ${\rm H^{+}}$, and ${\rm N^{+}}$.
The  ion fragments ${\rm NHD^{+}}$  (m/z=17) and ${\rm ND_2^{+}}$ (m/z=18) provided by the QMS in the vacuum chamber during the warming-up phase, can be added to the TPD signals of ionized NH$_3^{+}$ (m/z=17) and ${\rm NH_2D^{+}}$ (m/z=18) molecules, respectively.
This means that the TPD curves in the Figure~\ref{Fig3}a (m/z=17) peaking at 96~K and 150~K are the mixture of the ionized ${\rm NH_3^{+}}$, and the cracking pattern  ${\rm NHD^{+}}$ of the ionized ${\rm NHD_2^{+}}$ (m/z=19) and ${\rm NH_2D^{+}}$  (m/z=18).
Similarly, the TPD curves  in the Figure~\ref{Fig3}b (m/z=18),  peaking at 96~K and 150~K are the mixture of the ionized ${\rm NH_2D^{+}}$ molecules, and the cracking pattern ${\rm ND_2^{+}}$ of ionized ${\rm NHD_2^{+}}$ (m/z=19) and ${\rm ND_3^{+}}$  (m/z=20) of the deuterated molecules.

In our experimental conditions, the electron's energy of the QMS ion source is 32~eV. With this energy, only 30~$\%$ of molecules desorbing from the surface are ionized in the head of the QMS. So we can not determine the precise contribution of species having the same mass m/z to the QMS data, but we can assume that most of ammonia molecules desorbing from the surface are not fragmented in the QMS head but only ionized.

As previously discussed in section \ref{NH3+D}, TPD peaks observed at 150~K in figure~\ref{Fig3}, panels b), c) and d) match well with the desorption of water impurities H$_2$O, HDO, and D$_2$O, respectively.
The observed deuterated ammonia species in the TPD spectra are likely to be formed by H-D substitution reaction between the impinging D atoms and the ammonia adsorbed on the oxidized graphite surface. We excluded any energetic particles (photons, electrons and ions) in the formation of the deuterated ammonia species. Previous control experiments realized in the laboratory exclude any possible interaction of the electrons with the surface. The energetic particles produced in the microwave plasma of D atoms inside the beam-line can not reach the cold surface of the sample during the D exposure phase, and dissociate therefore the adsorbed ${\rm NH_3}$ molecules and cause their deuteration.

In Figure~\ref{Fig3}a, the strong TPD peak at 96~K (in black line) has the behavior of multilayer desorption of ${\rm NH_3}$ ice, where ${\rm NH_3}$ is probably bound to adsorbed ${\rm NH_3}$ by hydrogen bonds. While the TPD peak (in black line) at 150~K corresponds to  ${\rm NH_3}$ molecules physisorbed on the sites of the oxidized HOPG surface and contaminants (OD, CD...).
In Figure~\ref{Fig3}b (m/z=18), the TPD peaks at 96~K  that increase with the increase of the exposure dose of D atoms, are likely to be attributed  to ${\rm NH_2D}$ molecules (m/z=18), formed from the reaction  (${\rm NH_3+D}$) on ammonia ice deposited on the surface. In the same Figure~\ref{Fig3}b, the growth of the TPD peaks at 150~K with D exposure doses seems to be coherent with the decrease of the TPD peaks at 150~K in the Figure~\ref{Fig3}a. The desorption peaks at 150~K Figure~\ref{Fig3}b are likely to be attributed to ${\rm NH_2D}$ formed from the reaction (${\rm NH_3+D}$) on the surface of the oxidized HOPG. The maximum of these TPD peaks shifts towards the lower temperature with the increase of the peak height, and the  ${\rm NH_2D}$ coverage on the surface.This means that the interaction ${\rm NH_3-NH_3}$ with D atoms leads to the formation of deuterated species ${\rm NHD_2}$, ${\rm NHD_2}$,  and ${\rm ND_3}$, desorbing into the gas phase at 96~K, as seen in the TPD curves of  Figure~\ref{Fig3}b, c, and d, respectively. While  the successive deuteration of adsorbed ${\rm NH_3}$ on the oxidized graphite surface, produces deuterated ammonia molecules on stronger bending sites,  which desorb from the surface at 150~K (see Figure~\ref{Fig3}b, c, and d).

The astrophysical group of Watanabe et al. \cite{2005Nagaoka} have demonstrated experimentally the efficient formation of the deuterated isotopologue species of methanol at low surface temperature (10~K) by the D atoms exposure on ${\rm CH_3OH}$ ice. The isotopic species were observed and detected by infrared spectroscopy during the exposure of the adsorbates at 10~K. Similarly to methanol molecules, we believed that the deuteration reaction ${\rm NH_3+D}$ proceeds during the exposure phase of ${\rm NH_3}$ and D atoms on the oxidized HOPG substrate at 10~K, thanks to the tunneling process. At 10~K, D atoms are mobile \citep{2008Matar} and can diffuse on the surface to react with solid ammonia molecules. However, since in our experiments the deuterated species of ammonia are detected by TPD measurements from 10~K to 200~K, it is possible that the formation of the deuterated species of ammonia proceeds during the warming-up phase of the sample, rather than during the exposure of the reactants on the surface at 10~K. This assumption for the deuteration of ammonia by D atoms at higher surface temperatures is not taken into account in our experiments.

\subsection{Kinetics of CH$_3$OH+D reaction}\label{discussion}

\begin{figure*}
\centering
\includegraphics[width=17cm]{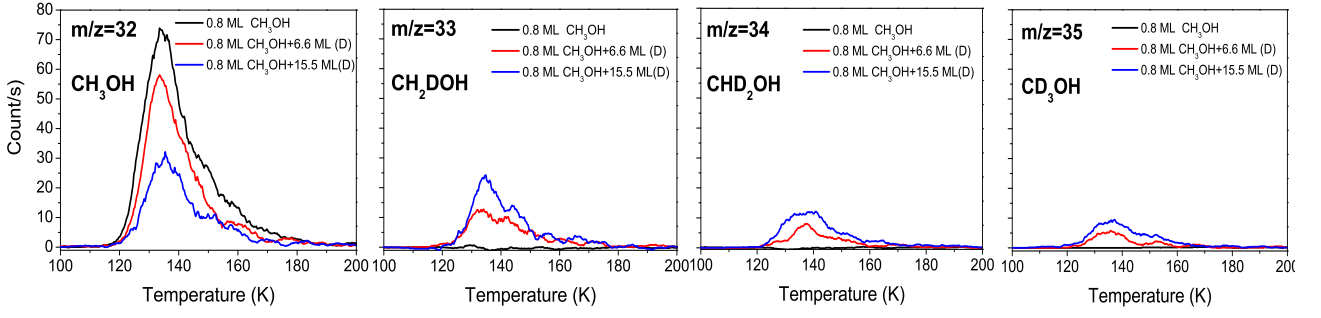}
\caption{TPD curves of ${\rm CH_3OH}$ (m/z=32), ${\rm CH_2DOH}$ (m/z=33), ${\rm CHD_2OH}$ (m/z=34) and ${\rm CD_3OH}$ (m/z=35) between
100~K and 200~K. Black curve: 0.8 ML of ${\rm CH_3OH}$ ice pre-deposited on the oxidized graphite surface held at 10~K; Red curve: after the exposure of 6.6~ML of D atoms at 10~K on 0.8 ML of ${\rm CH_3OH}$ ice; Blue curve: after the exposure of 15.5~ML of D atoms at 10~K on 0.8 ML of ${\rm CH_3OH}$ ice.} \label{Fig4}
\end{figure*}

In this section, we would like to compare the kinetic reaction ${\rm NH_3+D}$ to that of ${\rm CH_3OH+D}$ molecules in the sub-monolayer regime. We investigated similar deuteration experiments of solid ${\rm CH_3OH}$ by D atoms as for ammonia molecules. The experiments were performed under the same conditions: same low surface coverage ($\sim$ 0.8~ML), same
D atomic flux ${\rm \phi (D)=3.7\times10^{12}}{\rm molecules\cdot cm^{-2}\cdot s^{-1}}$, and same surface temperature (10~K).
Firstly, we deposited 0.8~ML of solid ${\rm CH_3OH}$ on the HOPG surface at 10~K, and then we added 6.6~ML of D atoms for the first experiment, and 15.5~ML of D atoms for the second one. After the D-addition phase, each film of ${\rm CH_3OH+D}$ was heated linearly from 10~K to 210~K using the same heating rate of 0.17~K$\cdot$s$^{-1}$. Figure~\ref{Fig4} shows the TPD desorption curves of ${\rm CH_3OH}$ (m/z=32), and the newly formed
isotopic species ${\rm CH_2DOH}$ (m/z=33), ${\rm CHD_2OH}$ (m/z=34) and ${\rm CD_3OH}$ (m/z=35) between 100~K and 200~K.
According to Nagaoka et al. \cite{2007Nagaoka} and Hiraoka et al. \cite{2005Hiraoka}, the H-abstraction of ${\rm CH_3OH}$ by D atoms
is likely to occur in the methyl ${\rm -CH_3}$ group rather than the hydroxyl ${\rm -OH}$ group of the ${\rm CH_3OH}$ (m/z=32) molecules.
We have thus attributed the TPD signals of masses m/z=33, m/z=34 and m/z=35 in the Figure~\ref{Fig4} to the newly formed deuterated species ${\rm CH_2DOH}$, ${\rm CHD_2OH}$, and ${\rm CD_3OH}$, respectively, which are deuterated in the methyl group.
The formation of the deuterated species in the hydroxyl group, such as ${\rm CH_3OD}$ (m/z=33), ${\rm CH_2DOD}$ (m/z=34),
and ${\rm CHD_2OD}$ (m/z=35), by the reaction system (${\rm CH_3OH+D}$) is expected to be negligible in this work. However, these deuterated methanol species in the hydroxyl group can be formed by D-H isotopic exchange between the species (${\rm CH_3OH}$, ${\rm CH_2DOH}$, and ${\rm CHD_2OH}$) with the deuterated ${\rm D_2O}$ water ice contaminants, during the transition phase from the amorphous to the crystalline state of the water ice at 120~K \citep{2009Ratajczak}.

\section{Analysis}\label{model and discussion}

\subsection{Rate equations of NH$_3$+D system reactions}\label{rate constant NH3}

We suggested that the reaction between ${\rm NH_3}$ and D atoms on the oxidized graphite surface held at 10~K, proceeds through direct Hydrogen-Deuterium substitution process by H-abstraction and D-addition mechanism, as proposed by
Nagaoka et al.~\citep{2005Nagaoka,2006Nagaoka} for ${\rm H_2CO~+~D}$ and ${\rm CH_3OH~+~D}$ reactions. In fact, the direct H-D substitution reaction (\ref{Eq2a}) leading to the formation of ${\rm NH_2D}$ species is
slightly exothermic with a formation enthalpy $\Delta${\rm H$^{0}$}= -${\rm 781.8~K}$.
\begin{equation}\label{Eq2a}
{\rm NH_3}~+~{\rm D}~{\longrightarrow}~{\rm NH_2D} + {\rm H},
\end{equation}
In the case of the H-abstraction and D-addition mechanism of ${\rm NH_3}$, the indirect H-D substitution process is described by the
following reactions (\ref{Eq2}) and (\ref{Eq3}).
 \begin{equation}\label{Eq2}
{\rm NH_3~+~D}~{\longrightarrow}~{\rm NH_2~+~HD},
\end{equation}
\begin{equation}\label{Eq3}
{\rm NH_2~+~D}~{\longrightarrow}~{\rm NH_2D},
\end{equation}

The first H-atom abstraction reaction~(\ref{Eq2}) of ${\rm NH_3}$ molecule by D atom leads to the formation of HD molecule and the ${\rm NH_2}$ radical. This reaction is endothermic with a reaction enthalpy of
${\rm \Delta H^{0}}$=${\rm +1527.5~K}$, and
needs an excess thermal energy to be produced. While the second D-addition reaction~(\ref{Eq3}) leading to the formation of the first
isotopologue ${\rm NH_2D}$ is exothermic with higher heat of formation $\Delta$${\rm H^{0}}$ = ${\rm-54480~K}$.
All the standard reaction enthalpies involving ammonia species and D atoms are provided by NIST database \citep{Nist} .

The same endothermic behavior takes place in the H-abstraction reactions~(\ref{Eq4}) and (\ref{Eq6}) of ${\rm NH_2D}$ and ${\rm NHD_2}$
species by D atoms, respectively.
\begin{equation}\label{Eq4}
{\rm NH_2D+D}~{\longrightarrow}~{\rm NHD+HD}       ~~~~({\rm \Delta{\rm H^{0}=+1455.3~K}})
\end{equation}
\begin{equation}\label{Eq4a}
{\rm NHD+D}~{\longrightarrow}~{\rm NH_2D}
\end{equation}
and
\begin{equation}\label{Eq6}
{\rm NHD_2+D}~{\longrightarrow}~{\rm ND_2+HD}       ~~~~({\rm \Delta{\rm H^{0}=+1527.5~K}})
\end{equation}
\begin{equation}\label{Eq6a}
{\rm ND_2+D}~{\longrightarrow}~{\rm ND_3}
\end{equation}

In order to fit the TPD experimental data of ${\rm NH_3}$, ${\rm NH_2D}$, ${\rm NHD_2}$ and ${\rm ND_3}$
species, given in the Figure~\ref{Fig6}, we used a kinetic model described by the following exothermic system reactions
(\ref{Eq8}-\ref{Eq10}) for the three direct H-D substitution reactions.

\begin{equation}\label{Eq8}
{\rm NH_3}+{\rm D} {\longrightarrow}{\rm NH_2D+H}~~~~(\Delta{\rm H^{0}}=-{\rm 781.8~K})
\end{equation}
\begin{equation}\label{Eq9}
{\rm NH_2D}+{\rm D}{\longrightarrow}{\rm NHD_2+H}~~~~(\Delta{\rm H^{0}}=-{\rm 938.1~K})
\end{equation}
\begin{equation}\label{Eq10}
{\rm NHD_2}+{\rm D}{\longrightarrow}{\rm ND_3+H}~~~~(\Delta{\rm H^{0}}=-{\rm 1131~K})
\end{equation}
These reactions are in competition with the exothermic D+D surface reaction leading to the formation of ${\rm D_2}$ molecules.
\begin{equation}\label{Eq11}
{\rm D}+{\rm D}{\longrightarrow}{\rm D_2}~~~~(\Delta{\rm H^{0}}=-{\rm 52440~K})
\end{equation}

Our model includes both Eley-Rideal (ER) and Langmuir-Hinshelwood (LH) mechanisms for the reactions of D atom either with another
D atom on the surface or with an ammonia species already adsorbed on the surface at 10~K. The Eley-Rideal mechanism occurs when one of
the species already adsorbed on the surface promptly reacts with a particle coming from the gas phase, before being adsorbed on the
surface. The Langmuir-Hinshelwood mechanism describes the formation of molecules on the surface when two adsorbed reaction-partners diffuse on the surface. D atoms are thermalized with the surface and they react with ammonia molecules thanks to surface diffusion. The ER mechanism is independent of the temperature of the surface ${\rm T_s}$, while LH mechanism is very sensitive to ${\rm T_s}$ since it depends on diffusion coefficients.
Moreover, LH is more efficient than ER mechanism at low surface coverage~\citep{2013Minissale}.
In our experiment, a D atom coming from the gas phase can hit an ammonia species already adsorbed on the surface, react and form a newly isotopic species of ammonia through ER mechanism. If the adsorbed D atom does not react through ER mechanism, it can diffuse on the surface from one site to a neighboring one. The diffused D atom can react either with another D atom on the surface to form ${\rm D_2}$ molecule, or with an adsorbed NH$_3$, NH$_2$D or NHD$_2$ molecules to form NHD$_2$, NH$_2$D, or ND$_3$ species, respectively, through the LH mechanism. All species, except D atoms, are not mobile on the surface at 10~K.

\subsubsection{Kinetic model}\label{model}

The model used to fit our experimental data is very similar to the one described by Minissale et al.~\citep{2013Minissale,2015Minissale}.
It is composed of six differential equations, one for each of the species considered:
D atoms, coming exclusively from the beam; NH$_3$ molecules, deposited on the surface; NH$_2$D, NHD$_2$, and ND$_3$, formed on
the surface; and finally D$_2$, coming both from the beam and formed on the surface. Each differential equation is composed of positive
and negative terms, indicating respectively an increase (i.e. species arriving from the gas phase or formed on the surface), or a
decrease (i.e. species reacting on the surface) in the surface coverage of the species.
The terms involving the ER and LH mechanisms are independent of one another, thus we are able to determine the amount of a
species formed (or consumed) via ER or LH mechanism.
Below, we present the list of differential equations governing the NH$_3$ deuteration:

\begin{align}\label{Eq12}
\frac {\rm d[{\rm D}]}{\rm {dt}}=&{\rm \phi_D} \biggl ({\rm 1-2 p_{1ER}[D]-p_{2ER}[NH_3]}-\nonumber \\
                       & {\rm p_{3ER}[{\rm NH_2D}]-p_{4ER}[NHD_2]}\biggr)-\nonumber \\
                       &{\rm k_{diff}}\biggl ({\rm 4p_{1LH}[D][D]-p_{2LH}[D][NH_3]}-\nonumber\\
                       &{\rm p_{3LH}[D][NH_2D]-p_{4LH}[D][NHD_2]}\biggr)
\end{align}
\begin{align}\label{Eq13}
\frac {\rm d[{\rm D_2}]}{\rm {dt}}=&{\rm \phi_D{_2}+2\phi_D (1-e_1)}{\rm p_{1ER}~[D]}+\nonumber \\
                         &2(1-e_1) {\rm k_{diff}}\cdot {\rm p_{1LH}[D][D]}
\end{align}
\begin{align}\label{Eq14}
\frac {\rm d[{NH_3}]}{\rm dt}=& -{\rm \phi_D} {\rm p_{2ER}}[{\rm NH_3}]-{\rm k_{diff}}\cdot {\rm p_{2LH}[D]}[{\rm NH_3}]
\end{align}

\begin{align}\label{Eq15}
\frac {\rm d[{NH_2D}]}{\rm dt}=&{\rm \phi_D} \biggl({\rm p_{2ER}}[{\rm NH_3}]- {\rm p_{3ER}}[{\rm NH_2D}]\biggr)+\nonumber\\
&{\rm k_{diff}}\biggl({\rm p_{2LH}} {\rm [D][NH_3]}-{\rm p_{3LH}}[{\rm D}][{\rm {NH_2D}}]\biggr)
\end{align}

\begin{align}\label{Eq16}
\frac {\rm d[{\rm NHD_2}]}{\rm dt}=&{\rm \phi_D} \biggl({\rm p_{3ER}}[{\rm NH_2D}]- {\rm p_{4ER}} [{\rm NHD_2}]\biggr)+\nonumber\\
&{\rm k_{diff}} \biggl({\rm p_{3LH}}[{\rm D}][{\rm NH_2D}]-{\rm p_{4LH}}[{\rm D}][{\rm NHD_2}]\biggr)
\end{align}
\begin{align}\label{Eq17}
\frac {\rm d[{ND_3}]}{\rm dt}={\rm \phi_D} \cdot {\rm p_{4ER}}[{\rm {NHD_2}}]+{\rm k_{diff}}\cdot {\rm p_{4LH}}[{\rm D}][\rm {NHD_2}]
\end{align}

The [${\rm D}$], [${\rm D_2}$], [${\rm NH_3}$], [${\rm NH_2D}$], [${\rm NHD_2}$], and [${\rm ND_3}$] quantities are the surface coverages of species. [X] is dimensionless and represents the percentage of surface covered with the X species. For each species [X] ranging between 0 and 1. This condition is true for all species except for D$_2$, whose surface coverage can be bigger than one. We stress that it does not represent a problem to evaluate activation barrier, since D$_2$ is an inert species and does not have any effect on kinetics of reactions. The initial reaction conditions at t=0 simulate the experimental conditions: $[\rm {NH_3}](t=0)$=~0.8 and ${\rm [NH_2D]=[NHD_2]=[ND_3]=0}$ for t=0. Furthermore, we impose that at any time:
\begin{align}
[\rm {NH_3}](t)+[\rm {NH_2D}](t)+[\rm {NHD_2}](t)+[\rm {ND_3}](t) \nonumber \\
=[\rm {NH_3}](t=0)
\end{align}
Dimensionless surface coverage is then converted in ML (or ${\rm molecule\cdot cm^{-2}}$) by multiplying [X] for the amount of adsorption sites of our surface ($10^{15}$~${\rm sites \cdot cm^{-2}})$ and compared with experimental results.
${\rm \phi_{X}}$ represents the part of surface covered per second by the X species coming from the gas phase.We know that in our experimental conditions the total number density of the impinging D-atoms coming from the gas phase and hitting the surface is given by the flux of D atoms in the beam-line: ${\rm \phi_{D}}$~=~${\rm 3.7\times10^{12}~atoms \cdot cm^{-2} \cdot s^{-1}}$. If we consider again that a surface contains 10$^{15}$~${\rm sites \cdot cm^{-2}}$, the flux of D atoms landing the surface is
${\rm \phi_{D}}$=${\rm 3.7\times10^{-3}~s^{-1}}$.

The terms concerning the chemical desorption of ammonia species ${\rm NH_2D}, {\rm NHD_2}$, and ${\rm ND_3}$ (formed by the reaction with D atoms) are not considered in the model, since the thermal desorption of these considered species is negligible at 10~K.
Despite the various heats of formation of ${\rm NH_3}$ isotopologues ($\Delta$${\rm H}$ less than 1100~K), no desorption of the newly formed ammonia species has been observed experimentally at 10~K from the graphite surface. This is because the local heats of formation of these species through the exothermic reactions (\ref{Eq8}-\ref{Eq10}) do not exceed the desorption energy of these ammonia species (${\rm E_{des}}$=2300~K)~\cite{2005Bolina}. So once the isotopologue ammonia species are formed, they stay in the solid phase on the graphite surface at 10~K, because their binding energy of about 2300~K is higher than the excess energies of formation.
The non chemical desorption of the corresponding molecules at 10~K is also confirmed experimentally by using the DED (During Exposure Desorption) method~\citep{2013Dulieu}, which consists of monitoring with the QMS placed in front of the sample, the signal of the deuterated molecules released into the gas phase during the deposition phase.
However, the parameter $e_1$ characterizing the prompt desorption of some ${\rm D_2}$ molecules, upon formation on the surface at 10~K, through the very exothermic reaction (\ref{Eq11}), is expected to be non negligible.
We point out that this term ${\rm e_1}$ does not influence ammonia species surface coverage.
In fact, as we have already said ${\rm D_2}$ is a non-reactive species and it cannot consume neither D atoms nor ${\rm NH_xD_y}$ species on the surface. However, it has been demonstrated by Amiaud et al. \cite{2007Amiaud} that the presence of D$_2$ molecules already adsorbed on the water ice increases the recombination efficiency of D atoms through the barrierless ${\rm D+D {\longrightarrow} D_2}$ reaction, by increasing the sticking coefficient of the deuterium atoms on the surface. This behaviour may have an important impact in the deuteration experiments, since the presence of condensed inert D$_2$ species may separate D and ${\rm NH_3}$, resulting in the decrease of the recombination efficiency of D atoms with adsorbed ${\rm NH_3}$, and the reduction of the H-D substitution reaction.

${\rm p_{1ER}}$ and ${\rm p_{1LH}}$ parameters are the reaction probabilities (dimensionless) of the D+D surface reaction (\ref{Eq11}), and we fixed to one their values.
Similarly, ${\rm p_{2ER}}$, ${\rm p_{3ER}}$, ${\rm p_{4ER}}$, and ${\rm p_{2LH}}$, ${\rm p_{3LH}}$ and ${\rm p_{4LH}}$ parameters represent the probabilities of the reactions (\ref{Eq8}-\ref{Eq10}) to be occurred via ER and LH mechanisms, respectively. Ammonia and D atoms species can react together by overcoming the activation barrier (Arrhenius term), or by crossing the barrier (tunneling term) as expressed by the following equations (\ref{Eq18}) and (\ref{Eq19}):

\begin{equation}\label{Eq18}
{\rm p_{iER}=e^{-Ea_i/(k_{B}\times T_{eff})} + e^{-2~Z_r \times \sqrt{2~\mu\times Ea_i\times k_{B}/h}}}
\end{equation}
and
\begin{equation}\label{Eq19}
{\rm p_{iLH}=e^{-Ea_i/(k_{B} \times T_{s})} + e^{-2~Z_r \times \sqrt{2~\mu\times Ea_i\times k_{B}/h}}}
\end{equation}

Where ${\rm k_B}$ is the Boltzmann constant, $h$ the Planck constant, ${\rm Ea_i}$ (i=2-4) are the activation energy barriers of the
reactions (\ref{Eq8}-\ref{Eq10}), ${\rm Z_r}$ is the width of the (rectangular) activation barrier, and ${\rm \mu}$ is the tunneling mass which is described by the reduced mass of the system involved in bi-molecular atom transfer reaction. This tunneling mass is defined as:
\begin{equation}
{\rm \mu=\frac{m_{NH_{x}D_{y}}\times m_D}{m_{NH_{x}D_{y}}+m_D}},~ \text{with x,y=0-3 and x+y=3}
\end{equation}
${\rm T_{eff}}$ is the effective temperature of the reaction between ${\rm NH_3}$, ${\rm NH_2D}$, ${\rm NHD_2}$ and D atoms given by:
\begin{equation}
{\rm T_{eff}= \mu(\frac{T_{solid}}{m_{NH_{3}}}+\frac{T_{gas}}{m_D})=314~K}
\end{equation}
${\rm T_{s}}$ (=10~K) is the surface temperature.
The parameter ${\rm k_{diff}}$ is the diffusion coefficient of D atoms between sites on the surface. It represents the amount of surface sites scanned in one second by D atoms.
It is defined by the following equation (\ref{Eq20})
\begin{align}\label{Eq20}
{\rm k_{diff}}=&{\rm \nu}~\biggl({\rm e^{-E_{diff}/k_{B}\times T_{s}}}+{\rm e^{-2~Z_d \times \sqrt{2~\mu \times E_{diff}\times k_{B}/h}}}\biggr)
\end{align}

Where ${\rm \nu=10^{12}~s^{-1}}$ is the attempt frequency for overcoming the diffusion barrier of D atoms and ${\rm E_{diff}}$ is the energy barrier for the diffusion of D atoms on cold surfaces held at 10~K. Bonfant et al. \cite{2007Bonfant} have reported an extremely low diffusion barriers of 4~meV for hydrogen atoms on graphite surface, meaning that hydrogen atoms physisorbed on graphite is highly mobile at low surface temperatures. However, for irregular surfaces such as ASW, the diffusion energy barrier of D atoms does not have a single value but follows a distribution, because there are several potential sites of different depths. Since our substrate used in our experiments is composed of an oxidized HOPG mixed with ASW ice deposits, the diffusion energy value of D atoms used in this model is that estimated on ASW ice for low surface coverage, ${\rm E_{diff}=(22\pm2)~meV~or~(255~\pm22)~K}$ \citep{2008Matar}. Even if this diffusion energy is different from that calculated on graphite surface, its high value does not affect the modeling results. The parameter
${\rm Z_d}$ is the width of the (rectangular) diffusion barrier. We have fixed the width of the diffusion barrier of D atoms (${\rm
Z_d}$) to 1~$\textup{\AA}$, a value commonly used to describe H or D atoms diffusion on the surface.
In our kinetic model, the term of the tunneling probability ${\rm e^{-2~Z_r \sqrt{2~\mu~Ea_i~k_B/h}}}$ for crossing the rectangular activation barrier depends on the tunneling mass of the reaction. This tunneling mass is described by the reduced mass $\mu$ of the system involving ammonia species ${\rm NH_{x}D_{y}}$ and D atom with x,y=0-3 and x+y=3. The value of $\mu$ for each reaction system is equal to 1.8 amu, and is close to the mass of the deuterium particle D (m/z=2 amu). This means that for direct D-H exchange reaction between D atom and ${\rm NH_3}$ molecules, the D atom is considered to be the tunneling particle that conduct the system ${\rm NH_3+D}$ to across the rectangular barrier through quantum tunneling.
According to Hidaka et al. \cite{2009hidaka}, the tunneling mass significantly depends on the reaction mechanism. For the addition reaction (${\rm AX+B~\rightarrow~AXB}$), the tunneling mass in the reaction coordinate is simply described by the reduced mass of the two-body system. However, for the abstraction reaction (${\rm AX+B~\rightarrow~A+BX}$), which involves three free particles in the reaction system, the tunneling mass is described by the effective mass, defined in the papers of Hidaka et al. \cite{2009hidaka}. For the direct H-D exchange reaction (\ref{Eq8}) between ${\rm NH_3}$ and D atoms, the description of the tunneling mass is not straightforward according to the reference ~\cite{2009hidaka}. However, if we assume that the H-D substitution reaction (\ref{Eq8}) occurs via an intermediate ${\rm NH_3D}$, having a tetrahedral geometry as demonstrated by ab-initio calculations~\citep{2005Moyano}, we can thus apply the reduced mass $\mu$ to describe the tunneling mass of the addition (\ref{Eq21}) reaction.

\begin{equation}\label{Eq21}
{\rm NH_3+D {\longrightarrow} NH_3D},
\end{equation}

\subsubsection{Activation energy barriers of the reactions}

In our kinetic model, we have four free parameters: the activation barriers (E$_{a2}$, E$_{a3}$, E$_{a4}$) for reaction (reaction
barrier) and the width of the reaction barrier ${\rm Z_r}$. This last can be constrained between 0.7 and 0.9~$\textup{\AA}$. Actually, solid-state chemistry at low temperatures should be dominated by quantum tunneling, according to Harmony~\citep{1971Harmony} and
Goldanskii~\citep{1976Goldanskii}. In particular, H-abstraction and D-substitution should be ruled by tunneling, as shown by Goumans et al. ~\citep{2011Goumans} in the case of CH$_3$OH deuteration. The reaction NH$_3$~+~D has been studied experimentally very long time ago in the gas phase by Kurylo et al. \cite{1969Kurylo} over the temperature range 423-741~K. These authors found that the H-D exchange between NH$_3$ and D may proceed through an intermediate NH$_3$D following the reaction ${\rm NH_3+D}$ ${\rightarrow}$ [${\rm NH_3D}$] ${\rightarrow}$ ${\rm NH_2D+H}$. As mentioned previously, the ${\rm NH_3+D}$ system reaction has been also studied theoretically \cite{2005Moyano} using ab-initio interpolated potential energy surface calculations. In both papers \citep{2005Moyano,1969Kurylo}, the activation barrier of the H-D exchange reaction (\ref{Eq8}) is reported to be ${\rm E_a=11~kcal/mol}$ or 5540~K.
However, some works (i.e. Bell~\citep{1980Bell} and references therein, Chapter 6: Tunneling in molecular spectra, the inversion of ammonia and related processes, page 153) argue that in the case of ammonia inversion, tunneling should be the dominant process, with a typical width of the reaction barrier (${\rm Z_r}$) is 0.7-0.8~$\textup{\AA}$. This reaction is considered the prototype of processes involving tunneling in a symmetrical (or quasi-symmetrical) potential energy curve. To the best of our knowledges no experimental and theoretical works have deal with the width of ammonia deuteration barrier. Thus we have used a value for width of deuteration similar to that of ammonia inversion, aware that the two reactions involve not identical chemical processes. For the sake of simplicity, we have used a common ${\rm Z_r}$ value for the deuteration reactions (\ref{Eq8}-\ref{Eq10}), instead of a value for each reaction. We suggest that tunneling is necessary for ammonia deuteration (in analogy with methanol deuteration). For this reason we use quantum tunneling in our model but we point out that our simple formulation of tunneling is useful only for a qualitative evaluation of our experimental results. Quantum tunneling refers to the quantum mechanical phenomenon where a particle tunnels through a barrier that it classically could not surmount. Quantum tunneling is known to be an important process for molecular synthesis on interstellar grains at very low temperatures \citep{2008Watanabe}. A detailed description of tunneling falls outside the scope of this work.

\begin{figure}
\centering
\includegraphics[width=8cm]{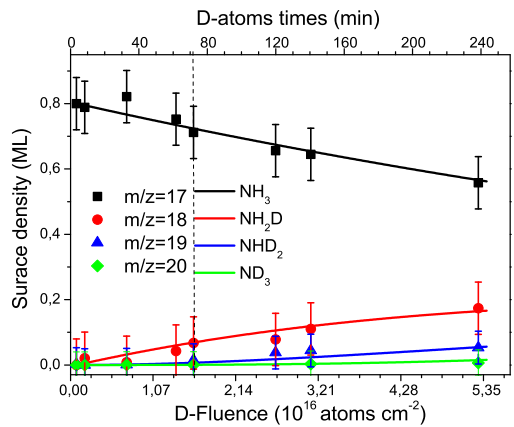}
\caption{Kinetic evolutions of ${\rm NH_3}$, ${\rm NH_2D}, {\rm NHD_2}$, and ${\rm ND_3}$ species present on the surface
as a function of D atoms exposure doses, and D fluences on 0.8~ML coverage of solid ammonia already deposited on the oxidized graphite substrate at 10~K. Black squares, red circles, blue triangles, green diamonds are the TPD experimental data
of ${\rm NH_3}$, ${\rm NH_2D}$, ${\rm NHD_2}$, and ${\rm ND_3}$, respectively. Solid lines are the fits
obtained from the model. The uncertainties are given by the errors bars.}
\label{Fig5}
\end{figure}

Figure~\ref{Fig5} shows the surface densities of NH$_3$, ${\rm NH_2D}$, NHD$_2$ and ND$_3$ species as a function of D atoms Fluences.
These surface densities, expressed in fraction of monolayer (ML) are the normalized integrated areas below the TPD curves of NH$_3$, ${\rm NH_2D}$, NHD$_2$ and ND$_3$ peaked at 97~K for each D atom fluence, with respect to the TPD integrated area of NH$_3$ for one monolayer coverage.
As previously explained  in the section \ref {Kinetic NH3+D}, we assumed that all the deuterated ammonia species are ionized by electron impact in the ion source of the QMS during the TPD experiments. However, it has been reported by Rejoub et al. \cite{2011Rejoub} that the ionization cross-section of light ${\rm NH_3}$ molecules is twice larger than that of  ${\rm ND_3}$, meaning that the cross-sections for formation of ion fragments from heavy deuterated molecules, are much smaller than those from NH$_3$. Because we neglected the contribution of the cracking patterns in the TPD experiments, we did not considered the different ionization cross-section values of  the deuterated species in the measurements of the areas from TPD profiles.
As shown in the Figure~\ref{Fig5}, there is a good correlation between the experimental data and the fits obtained by the model for the exponential decay of NH$_3$, and the growth of ${\rm NH_2D}$, NHD$_2$ and ND$_3$ species on the surface when increasing the amount of D atoms on the surface. The plots of the Figure~\ref{Fig5} show that for the higher D-irradiation time of 240 minutes, and the higher D-fluence of ${\rm 5.34 \times 10^{16}~atoms \cdot cm^{-2}}$, about 20~\% of the adsorbed NH$_3$ molecules are mainly deuterated into ${\rm NH_2D}$ and ${\rm NHD_2}$ species with traces of ${\rm ND_3}$. The formation yields of the single, the double and the triple deuterated ammonia species are approximately ($14~\%$), ($5~\%$), and ($1~\%$), respectively.

Thanks to our model, we have tested different scenario: we used three values (150~K, 250~K, 350~K) for the diffusion barrier ${\rm E_{diff}}$ of D atoms, and for each value, we have varied ${\rm Z_r}$ from 0.7 to 0.9~$\textup{\AA}$ (step of 0.01 $\textup{\AA}$).
In the case of ${\rm E_{diff}= 250~K}$, the activation energy barriers of the successive H-D substitution reactions of ammonia species by
D atoms are found to be ${\rm Ea_2=(1840\pm270)}$~K for the reaction~(\ref{Eq8}), ${\rm Ea_3=(1690\pm245)}$~K for the reaction~(\ref{Eq9}), and ${\rm Ea_4}$ = ${\rm (1670\pm230)}$~K for the reaction~(\ref{Eq10}).
In the Table \ref{table1}, we list the width and the energy activation barrier for the H-D substitution reactions of
${\rm NH_2D}$, NHD$_2$ and ND$_3$ species. The listed values of the activation energies ${\rm Ea_i}$ minimize the $\chi^2$ value of our fit with respect to our experimental data.
The statistical parameter $\chi^2$ is obtained for each set of parameters by using the following formula:
\begin{equation}
{\rm \chi^2=\Sigma_{ML, mol} [S_t(ML, mol)-S_e(ML, mol)]^2/ S_e(ML, mol)}
\end{equation}
Where ${\rm S_t(ML, mol)}$ and ${\rm S_e(ML, mol)}$ are respectively the theoretical and experimental surface density for each molecule at a certain D-fluence.

\begin{figure}
\centering
\includegraphics[width=8cm]{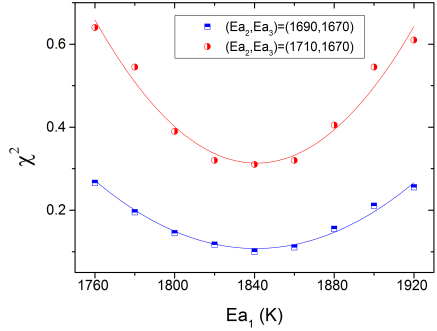}
\caption{The curves minimizing the $\chi^2$ value between the kinetic model and the experimental measurements for the reaction between ${\rm NH_3}$ and D atoms, by setting a couple of activation energies barriers (${\rm Ea_2,~Ea_3}$) for the deuteration reactions~(\ref{Eq8}) and (\ref{Eq9}) and varying the value of third one (${\rm Ea_4}$) for the reaction~(\ref{Eq10}). ${\rm E_{diff}}$ and Z$_r$ are respectively 250~K and 0.83~$\textup{\AA}$. }
\label{Fig6}
\end{figure}

In order to have a good correlation between the model and the experiments, Figure~\ref{Fig6} shows how we can minimize the $\chi^2$ value by setting a couple of activation energies values (${\rm Ea_2,~Ea_3}$) for the deuteration reactions~(\ref{Eq8}) and (\ref{Eq9}) and varying only the value of the third energy ${\rm Ea_4}$ for the reaction~(\ref{Eq10}).
Our activation energy barriers for the direct H-D substitution reactions (\ref{Eq8}-\ref{Eq10}) between ammonia species and D atoms (see Table~\ref{table1}) are smaller than the activation energy barrier ${\rm E_a=5540~K}$ reported by the two references \cite{1969Kurylo, 2005Moyano}, both in gas and solid phases. The low values of the activation energies obtained in this work can be explained by the catalytic effect of the ASW ice~+~oxidized HOPG on the deuteration reaction ${\rm NH_3+D}$. This substrate favors the diffusion of D atoms, and increases therefore the reactivity between ${\rm NH_3}$ molecules and D atoms on the surface.

As shown in the Table~\ref{table1}, the width ${\rm Z_r}$ and the energy of the activation barriers ${\rm E_a}$ depend
on the diffusion energy of D atoms on the surface ${\rm E_{diff}}$. One can note that the higher the diffusion energy (${\rm E_{diff}}$) of D atoms, the lower the width (${\rm Z_r}$) of the energy barriers. The diffusion of D atoms on the cold surface increases the probability of the H-D substitution reactions of ammonia in the solid phase.
Table \ref{table1} also shows that for each ${\rm E_{diff}}$, the value of the activation energy barrier is always high for the first deuteration reaction ${\rm NH_3+D}$, and then decreases by almost $10~\%$ for the second ${\rm NH_2D+D}$ and the third ${\rm NHD_2+D}$ deuteration reactions. Our activation energy barriers for the deuteration reaction ${\rm NH_3+D}$ in the solid phase is much lower than the value (${\rm 46~kJ\cdot mol^{-1}}$) given in the gas phase \citep{2005Moyano,1969Kurylo}. This large difference can be explained by the catalytic effect of substrate composed of oxidized graphite and ASW ices deposits.

\begin{table}
\centering \caption{The width Z$_r$ and the height of the energy barriers E$_a$, expressed in ($\textup{\AA}$) and in kelvin (K), respectively, of the successive H-D substitution reactions of ${\rm NH_3}$ molecules by D atoms on the oxidized, partly ASW covered graphite surface at 10~K, for a fixed value of D-atom diffusion energy ${\rm E_{diff}}$. The minimum $\chi^2$ value of the fits is found to vary between 0.3 and 0.1.}\label{table1}
\begin{tabular}{c|c|c|ccc}
\hline\hline
Reactions & ${\rm E_{diff}}$                                                        & Z$_r$ & E$_a$  \\
\hline
 units         &  K                                                            & $\textup{\AA}$    & K& \\
\hline
${\rm NH_3}$+D  $\overset{p_{2}}{\longrightarrow}$ ${\rm NH_2D}$+H &150 & 0.86  &   1950 $\pm$ 250& \\
                                                                   &250 & 0.83  &   1840 $\pm$ 270& \\
                                                                   &350 & 0.81  &   1750 $\pm$ 320& \\
\hline
${\rm NH_2D}$+D $\overset{p_{3}}{\longrightarrow}$ ${\rm NHD_2}$+H &150 & 0.86  &   1820  $\pm$ 220& \\
                                                                   &250 & 0.83  &   1690  $\pm$ 245&  \\
                                                                   &350 & 0.81  &   1610  $\pm$ 290&  \\
\hline
${\rm NHD_2}$+D $\overset{p_{4}}{\longrightarrow}$ ${\rm ND_3}$+H  &150 & 0.86  &   1800 $\pm$ 210&\\
                                                                   &250 & 0.83  &   1670  $\pm$ 230&  \\
                                                                   &350 & 0.81  &   1600  $\pm$ 250&  \\
\hline\hline
\end{tabular}
\end{table}

\subsection{Rate equations of CH$_3$OH + D system reactions }\label{barrier CH3OH}

As has been suggested by Nagaoka et al.~\citep{2007Nagaoka}, the deuteration of ${\rm CH_3OH}$, ${\rm CH_2DOH}$ and ${\rm CH_2DOH}$ species by D atoms on cold surfaces, occur through the successive H abstraction and D addition mechanism as follows:

\begin{equation}\label{Eq25}
{\rm CH_3OH~+~D}~{\longrightarrow}~{\rm CH_2OH +HD}
\end{equation}
\begin{equation}\label{Eq26}
{\rm CH_2OH~+~D}~{\longrightarrow}~{\rm CH_2DOH}
\end{equation}
\begin{equation}\label{Eq27}
{\rm CH_2DOH+D}~{\longrightarrow}~{\rm CHDOH +HD}
\end{equation}
\begin{equation}\label{Eq28}
{\rm CHDOH +D}~{\longrightarrow}~{\rm CHD_2OH}
\end{equation}
\begin{equation}\label{Eq29}
{\rm CHD_2OH +D}~{\longrightarrow}~{\rm CD_2OH+HD}
\end{equation}
\begin{equation}\label{Eq30}
{\rm CD_2OH~+~D}~{\longrightarrow}~{\rm CD_3OH}
\end{equation}

Where the H-abstraction reactions (\ref{Eq25}, \ref{Eq27} and \ref{Eq29}) are exothermic with small activation
barriers in comparison to the direct H-D substitution reactions. While the D-addition reactions (\ref{Eq26}, \ref{Eq28} and \ref{Eq30}) are exothermic with no barriers. In Hama et all's review~\citep{Hama2013}, the direct H-D substitution reaction
${\rm CH_3OH+D}$~${\longrightarrow}$~${\rm CH_2DOH +H}$ has a very large activation energy barrier of 169~kJ/mol (or 20330~K), in comparison to the following H-abstraction reaction
${\rm CH_3OH~+~D}$~${\longrightarrow}$~${\rm CH_2OH +HD}$, which has an activation energy of ${\rm 27~kJ\cdot mol^{-1}}$ (or 3250~K), estimated from gas phase calculations.

In this work, the successive deuteration reactions of methanol species by D atoms are described by the following simple reactions
(\ref{Eq31}-\ref{Eq33}).
\begin{equation}\label{Eq31}
{\rm CH_3OH+D} \overset{p'_{2}}{\longrightarrow}.....{\longrightarrow}{\rm CH_2DOH}
\end{equation}
\begin{equation}\label{Eq32}
{\rm CH_2DOH+D} \overset{p'_{3}}{\longrightarrow}.....{\longrightarrow}{\rm CHD_2OH}
\end{equation}
\begin{equation}\label{Eq33}
{\rm CH_2DOH+D} \overset{p'_{4}}{\longrightarrow}.....{\longrightarrow}{\rm CD_3OH}
\end{equation}

Where ${\rm p'_2, p'_3~and~p'_4}$ parameters are the reaction probabilities of the H-abstraction reactions (\ref{Eq25}, \ref{Eq27} and \ref{Eq29}) of ${\rm CH_3OH}$, ${\rm CH_2DOH}$ and ${\rm CH_2DOH}$ by D atoms, respectively.
Using the same kinetic model previously described for ${\rm NH_3+D}$ system reactions in solid phase, we have
estimated the activation energy barriers of the H-abstraction reactions (\ref{Eq25}, \ref{Eq27} and \ref{Eq29}) for ${\rm CH_3OH+D}$ system reactions.
Table~\ref{table2} summarizes the values of the width Z$_r$ and the activation energy barriers E$_a$ for each H-D substitution reaction of methanol species.
Our kinetic model provides an activation energy barrier E$_a$=(1450~$\pm$~210)~K or (12.1~$\pm$~1.7)~${\rm kJ\cdot mol^{-1}}$ for the first abstraction reaction (${\rm CH_3OH+D}$) given by the equation (\ref{Eq25}). This value is more than a factor of two smaller
than the activation energy value (${\rm 27~kJ\cdot mol^{-1}}$ or ${\rm 3250~K}$), reported by \cite{Hama2013} from theoretical estimations in the gas phase. Once again the catalytic role of the surface can explain the difference between gas phase and solid phase activation barriers.
The activation energy barriers of the successive deuteration reactions of methanol species
decease significantly with the increase of the diffusion energy ${\rm E_{diff}}$ of D atoms on the surface.
Nerveless, the values of the activation energy barriers for ${\rm CH_3OH+D}$ system reactions (see Table~\ref{table2})
are always smaller than those of the ${\rm NH_3+D}$ system reactions (see Table~\ref{table1}).

\begin{table}
\centering \caption{The width Z$_r$ and the height of the energy barriers E$_a$, expressed in ($\textup{\AA}$) and in kelvin (K), respectively, of the successive H-D substitution reactions of ${\rm CH_3OH}$ molecules by D atoms on the oxidized, partly ASW covered graphite surface at 10~K, for a fixed value of D-atom diffusion energy ${\rm E_{diff}}$. The $\chi^2$ value of the fit vary between 0.3 and 0.1. \label{table2}}
\begin{tabular}{c|c|c|ccc}
\hline\hline
Reactions & ${\rm E_{diff}}$                                                      & Z$_r$ & E$_a$  \\
\hline
 units          &  K                                                            &  $\textup{\AA}$   & K&  \\
\hline
${\rm CH_3OH}$+D  $\overset{p'_{2}}{\longrightarrow}$ ${\rm CH_2DOH}$+H &150 & 0.86  &   1450 $\pm$ 210& \\
                                                                        &250 & 0.85  &   1080 $\pm$ 180&  \\
                                                                        &350 & 0.84  &   860  $\pm$ 120& \\
\hline
${\rm CH_2DOH}$+D $\overset{p'_{3}}{\longrightarrow}$ ${\rm CHD_2OH}$+H &150 & 0.86  &   1330 $\pm$ 200&\\
                                                                        &250 & 0.85  &   990  $\pm$ 180&\\
                                                                        &350 & 0.84  &   770  $\pm$ 145&\\
\hline
${\rm CHD_2OH}$+D $\overset{p'_{4}}{\longrightarrow}$ ${\rm CD_3OH}$+H  &150 & 0.86  &   1300 $\pm$ 205&\\
                                                                        &250 & 0.85  &   980  $\pm$ 170&\\
                                                                        &350 & 0.84  &   780  $\pm$ 150&\\
\hline\hline
\end{tabular}
\end{table}

\begin{figure}
\centering
\includegraphics[width=8cm]{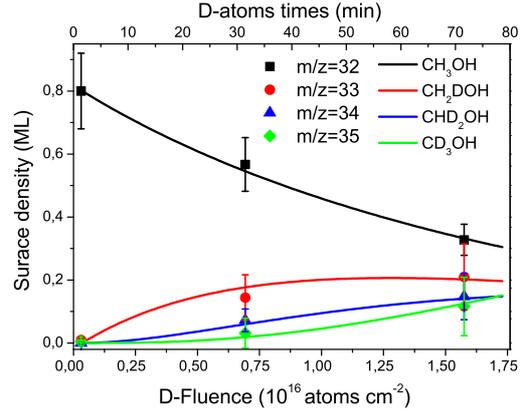}
\caption{The variation of the surface densities of ${\rm CH_3OH}$ (m/z=32), ${\rm CH_2DOH}$ (m/z=33), ${\rm CHD_2OH}$
(m/z=34) and ${\rm CD_3OH}$ (m/z=35) as a function of D atoms exposure times and D fluences on 0.8~ML coverage of ${\rm CH_3OH}$ film pre-deposited on the surface of an oxidized graphite substrate held at 10~K. Full squares, circles and triangles are the TPD experimental data. Solid lines are the fits obtained from the model. The uncertainties are given by the errors bars.} \label{Fig7}
\end{figure}

Figure~\ref{Fig7} shows the best fit of the data for the exponential decay of ${\rm CH_3OH}$,
and the increase of the surface densities of ${\rm CH_2DOH}$, ${\rm CHD_2OH}$ and ${\rm CD_3OH}$ with the increasing time and fluence of D atoms exposure on 0.8~ML of solid methanol ${\rm CH_3OH}$ pre-deposited on the oxidized HOPG surface.
We can note that after 70~minutes of D atoms addition, about ${\rm 0.44~ML}$ of the adsorbed ${\rm CH_3OH}$ molecules, are deuterated into three isotopologue species, with the formation yields of ($\sim22~\%$) for ${\rm CH_2DOH}$, ($\sim15~\%$) for ${\rm CHD_2OH}$, and ($\sim8~\%$) for ${\rm CD_3OH}$. The comparison with the previous kinetic results given in the Figure~\ref{Fig6}, shows that when 0.8~ML of solid NH$_3$ is irradiated with D atoms during the same exposure time of 70 minutes, only $10~\%$ of NH$_3$ molecules are deuterated into ${\rm NH_2D}$ ($\sim8~\%$) with traces of ${\rm NHD_2}$ ($\sim2~\%$) and ${\rm NHD_3}$ ($<1~\%$). This means that during 70 minutes of D atoms exposure phase, 0.44~ML (or 55~\%) of adsorbed ${\rm CH_3OH}$ molecules are consumed by D atoms, while only 0.08~ML (or 10~\%) of solid ammonia can be destructed by D atoms.
We defined the deuteration rate ${\rm \upsilon_X=\frac{d[X]}{dt}}$ of an adsorbed species X by D atoms, as the amount [X] of this species (in ${\rm molecules}$ ${\rm \cdot cm^{-2}}$), consumed per unit time (in ${\rm min}$) of D atoms exposed on the surface. The deuteration rate value of ${\rm CH_3OH}$ is estimated to ${\rm \upsilon_{CH3OH}}$  ${\rm\backsimeq 0.005\times10^{15}}$ ${\rm molecules}$ ${\rm \cdot cm^{-2}\cdot min^{-1}}$, while that of NH$_3$ species is found to be ${\rm \upsilon_{NH3}}$ ${\rm \backsimeq 0.001\times10^{15}}$ ${\rm molecules\cdot cm^{-2}\cdot min^{-1}}$, and can slightly decrease for extended irradiation up to 240 minutes. The relationship between the two deuteration rates is ${\rm \frac{\upsilon_{CH3OH}}{\upsilon_{NH3}}\simeq5}$, meaning that the deuteration rate of ${\rm CH_3OH}$ molecules by D atoms on cold and oxidized graphite HOPG surfaces with ASW ice deposits, is  five times higher than of NH$_3$.

\section{Discussion and conclusions}

In this work, we demonstrated experimentally the possible deuteration of ${\rm NH_3}$ molecules by D atoms on cold oxidized HOPG surface, partly covered with ASW ices. The deuteration experiments of solid ammonia were performed at low surface coverage and low temperature 10~K using mass spectroscopy and temperature programmed desorption (TPD).
The isotopologue ammonia species NH$_2$D, NHD$_2$ and ND$_3$ desorbing from the surface at 96~K and 150~K are  likely to be formed  by direct exothermic H-D substitution reactions between the adsorbed ammonia species on the surface and the impinging D atoms. A kinetic model taking into account the diffusion of D atoms on the surface provides the activation energy barriers of the deuteration reactions ${(\rm NH_3+D)}$ in the solid phase. We found that the energy barrier for the D-H exchange reaction ${(\rm NH_3+D)}$ is (1840~$\pm$~270~K) or (${\rm 15.4~\pm~2.5~kJ\cdot mol^{-1}}$), three  times lower than that predicted in the gas phase (5530~K or ${\rm 46~kJ\cdot mol^{-1}}$) \citep{1969Kurylo,2005Moyano}. Our results also show that the activation energy barrier for the first deuteration reaction ${(\rm NH_3+D)}$ of ammonia is almost two times higher than that of the abstraction reaction (\ref {Eq25}) of ${\rm CH_3OH}$  (1080~$\pm$~180~K) or (9.0~$\pm$~1.2~${\rm kJ\cdot mol^{-1}}$).
Our experimental results showed that the deuteration reaction (${\rm NH_3+D}$) occurs through quantum tunneling, and it is five orders of magnitude slower than methanol (${\rm CH_3OH}$) deuteration process.

If our laboratory experiments lead to the formation of the deuterated ammonia species in comparison to the previous experiments of Nagaoka et al. \citep{2005Nagaoka} and Fedoseev et al. \citep{2015Fedoseev}. This is because our experimental conditions are different from the other works, and help to overcome  the classical, the quantum tunneling, and the diffusion activation barriers of the reactions between ${\rm NH_3}$ and D atoms on the oxidized graphite surface. It seems that the main factors that enhance the deuteration reactions between ammonia and D atoms in our experiments are the low D flux and the low thickness of solid ${\rm NH_3}$. The effect of the ASW water ices contaminants on the formation of the observed deuterated ammonia species seems to be negligible.

By lowering significantly the D atoms-flux in this work with respect to the previous works of Fedoseev et al. \citep{2015Fedoseev} and Nagaoka et al. \citep{2005Nagaoka}, we increase the density of D atoms available to diffuse on the surface, and interact efficiently with  physisorbed ammonia species on the oxidized HOPG surface. However, the D-H exchange reaction between ${\rm NH_3}$ and D atoms is almost in competition with the barrier-less recombination ${\rm D}+{\rm D}{\longrightarrow}{\rm D_2}$ reaction. We notice that the used D-flux could be suitable to have H-D substitution reactions in the experiments of Fedoseev et al. \citep{2015Fedoseev} and Nagaoka et al. \citep{2005Nagaoka}, both for methanol  and ammonia, if they have reduced the thickness of their ices to a fraction of one monolayer. These authors did not try to reduce simultaneously the thickness of the ices and the flux of D atoms to study the effect of these two parameters on the the efficiency of the deuteration reactions at low temperatures.
Other factors related to a specific orientation of the graphitic surface, or a possible interaction of ammonia with the substrate by chemisorption process, at low surface temperature at 10~K, can induce the H-D exchange between ${\rm NH_3}$ and D atoms. Some experimental \citep{2013Yeh} and theoretical \citep{2012Tang}
works reported in the literature have demonstrated the possible dissociative chemisorption of NH$_3$ molecules on the oxidized graphite surface at the epoxy functional groups, created by the oxidation of the HOPG surface at room temperature \citep{2012Larciprete}. The dissociation of the adsorbed ammonia leads to the formation of ${\rm C-NH_2}$ radicals, which can react with deuterium atoms. If the chimisorption occurs for NH$_3$ in our experiments, CH$_3$OH molecules will be also chemisorbed on the oxidized graphite surface, and leads to the formation of the radical CH$_2$OH. It may be the case, but all the molecules in our experiments desorb at the physisorption temperatures, at 140~K for methanol, 96~K for ammonia species, and 150~K for traces of water ices. In addition, the energy barrier for the dissociative chemisorption of NH$_3$ on the oxidized HOPG surface has been predicted to be 97.90~${\rm kJ\cdot mol^{-1}}$ \citep{2012Tang}. This activation energy barrier is significantly high to be overcome, in comparison to the activation energy barriers found in this work (16 ${\rm kJ\cdot mol^{-1}}$), and makes chemisorption improbable in our experiments.

Our results for the deuteration of small molecules, such as NH$_3$ by surface chemistry are important in the fields of astronomy, astrochemistry and low-temperature physics.
The formation of the isotopologue ammonia species on cold grain surfaces can contribute to the D-enrichment of ammonia in the interstellar medium, and explain therefore the observed ratios of H-and D-ammonia bearing molecules in dark clouds. However, the amount of the NH$_3$ molecules expected to be deuterated in dense molecular clouds on a timescale of ${\rm 10^{5}-10^{6}}$ years is not enough important to reproduce the large gas-phase interstellar abundances of deuterated ammonia molecules.
In space, the low reaction probability of NH$_3$ molecules with D atoms on interstellar grain mantles results from the competitive surface reactions of D atoms with the accreted species (D, N, O, O$_2$...) from the gas phase, leading to the formation of D$_2$ molecules through D+D recombination \cite{2006Horneker}, NH$_2$D and ND$_3$ through D+N addition reactions \cite{2015Fedoseev}, and heavy D$_2$O water ices through D+O and D+O$_2$ chemical reactions \cite{2010Ioppolo,2013Dulieu, 2013Chaabouni}.

Observational studies towards cold regions (dark molecular clouds and dense cores L134N ) \citep{2000Roueff, 2000Saito,2002Van} show deuterated ammonia species in gas phase. In these cold environments of the interstellar medium, grain mantles are exposed to cosmic rays and UV irradiation fields originating from hot stars  These energetic UV photons may induce the desorption into the gas phase of the newly formed deuterated species NH$_2$D, NHD$_2$ and ND$_3$ on the grain surfaces by non thermal photo-desorption process \cite{1990Hartquist}. The desorption of the deuterated ammonia species in cold regions may also result from exothermic reactions occurring on the grain, in particular the formation of molecular hydrogen (${\rm H_2}$) \cite{1993Duley} by H bombardment. This reaction releases 4.5 eV of excess energy which can be transferred to the grain surface, and causes the local heating of the deuterated species. It has been demonstrated that such local heating can reorganize the local structure of the ice mantle \cite{2011Accolla}, although it can hardly induce indirect desorption \cite{2016Minissale}.

Because the abundance of ammonia (NH$_3$) in icy mantles is nearly 15~\% with respect to water (H$_2$O) ice
\citep{2000Gibb,2004Gibb}, it is interesting to study the efficiency of the deuteration reaction ${\rm NH_3+D}$ on amorphous solid water ASW ice surfaces, and explore the role of the ice grain chemistry in the interstellar deuterium fractionation of ammonia molecules.

\section{Acknowledgements}

The authors thank the editor and the anonymous reviewers for their valuable and useful comments. The authors also thank Dr Paola Caselli (Center for Astrochemical Studies, Max Planck Institute for Extraterrestrial Physics MPE) for the relevant discussions about the deuteration of ammonia in the interstellar medium.


\bibliographystyle{spphys}       
\bibliography{REF.NH3B}

\end{document}